\documentclass[a4paper,12pt]{article}
\pdfoutput=1
\usepackage{amsfonts,amsthm,amsmath,amssymb,graphicx}
\usepackage{cite}
\usepackage{braket}
\usepackage{amsmath}
\usepackage{url}
\usepackage{graphicx,color}

\usepackage[dvipsnames]{xcolor}
\usepackage[colorlinks=true,urlcolor=Bittersweet,linkcolor=Bittersweet,citecolor=Bittersweet]{hyperref}

\DeclareMathOperator\arcosh{arccosh}
\DeclareMathOperator\arctanh{arctanh}

\def\tr{{\mathrm{tr}}}

\def\sinh{{\mathrm{sinh}}}
\def\cosh{{\mathrm{cosh}}}
\def\tanh{{\mathrm{tanh}}}

 \def\frac#1#2{{#1\over #2}}

 \def\Re{{\rm Re}}
 
\def\be{\begin{equation}}
\def\ee{\end{equation}}
\def\bea{\begin{eqnarray}}
\def\eea{\end{eqnarray}}
\numberwithin{equation}{section}

\usepackage{hyperref}
\begin{document}


\begin{titlepage}

\thispagestyle{empty}

\begin{flushright}
YITP-18-111
\end{flushright}

\bigskip

\begin{center}
\noindent{{\large \textbf {Time Evolution of Complexity: \\
A Critique of Three Methods }}}\\
\vspace{1cm}
	
Tibra Ali$^{(a)}$\footnote{tali@perimeterinstitute.ca}, Arpan Bhattacharyya$^{(b)}$\footnote{bhattacharyya.arpan@yahoo.com},  
S. Shajidul Haque$^{(c)}$\footnote{shajid.haque@uwindsor.ca}, Eugene H. Kim$^{(c)}$\footnote{ehkim@uwindsor.ca} 
and\\ Nathan Moynihan$^{(d)}$\footnote{nathanmoynihan@gmail.com} \\ ~~~~\\

$ {}^{(a)}$ {\it Perimeter Institute, \\
\it 31 Caroline Street North\\
 Waterloo, Ontario, Canada, N2L 2Y5 \\
}	
        
$ {}^{(b)}$ {\it Center for Gravitational Physics, \\
\it Yukawa Institute for Theoretical Physics (YITP), Kyoto University, \\
\it Kitashirakawa Oiwakecho, Sakyo-ku, Kyoto 606-8502, Japan\\
}	

$ {}^{(c)}$ {\it Department of Physics, University of Windsor, \\
\it 401 Sunset Avenue\\
Windsor, Ontario, Canada, N9B 3P4\\
}	

$ {}^{(d)}$ {\it{The Laboratory for Quantum Gravity \& Strings,\\
	Department of Mathematics \& Applied Mathematics,\\
	University of Cape Town,\\
	Private Bag, Rondebosch, 7701, South Africa}\\ 
}
\vskip 2em
\end{center}

\begin{abstract}
\vspace{-0.75cm}

\end{abstract}
In this work, we propose a testing procedure to distinguish between the different approaches for computing complexity. Our test does not require a direct comparison between the approaches and thus avoids the issue of choice of gates, basis, etc. The proposed testing procedure employs the information-theoretic measures Loschmidt echo and Fidelity; the idea is to investigate the sensitivity of the complexity (derived from the different approaches) to the evolution of states. 
We discover that only circuit complexity obtained directly from the wave function is sensitive to time evolution, leaving us to claim that it surpasses the other approaches. 
We also demonstrate that circuit complexity displays a universal behaviour---the complexity is proportional to the number of distinct Hamiltonian evolutions that act on a reference state. 
Due to this fact, for a given number of Hamiltonians, we can always find the combination of states that provides the maximum complexity; consequently, other combinations involving a smaller number of evolutions will have less than maximum complexity and, hence, will have resources. Finally, we explore the evolution of complexity in non-local theories; we demonstrate the growth of complexity is sustained over a longer period of time as compared to a local theory.

\end{titlepage}
\newpage
\tableofcontents


\section{Introduction}

Recent progress in the fields of quantum information and condensed matter have shed light on the inner-workings of holographic duality (AdS/CFT duality). It is becoming evident that the entanglement entropy (EE) defined in the boundary conformal field theory (CFT) is related to the emergence of the bulk geometry; this relationship becomes more stimulating in the context of black hole physics \cite{RT,RTa, RT1,bookRT}\footnote{A massive literature exists in this context; the reader is encouraged to consult Ref.~\cite{bookRT} for a comprehensive list of references.}. In the black hole setting, an important question is, ``{\sl What is the appropriate observable which can probe physics behind the horizon?" } 
It was observed that although the EE saturates as the black hole thermalizes \cite{Hartman}, the size of the Einstein-Rosen bridge (of an eternal AdS black hole) continues to increase. Based on this observation, Susskind et al.~proposed that the quantity in the CFT that continues to increase after thermalization is the {\sl complexity} \cite{MS, Susskind1,Susskind2}. Two interesting proposals were made in the context of AdS/CFT \cite{MS, Susskind1,Susskind2}. The first is `complexity equals volume' (CV conjecture)---the volume is that of a maximal co-dimension-one bulk surface extending to the boundary of AdS space time, which can be chosen to asymptote to a specific time-slice where the boundary state resides. The second is `complexity equals action' (CA conjecture)---one evaluates the bulk action (with suitable boundary and counter terms to make the variational principle well defined) on the so-called Wheeler-DeWitt (WDW) patch. 
Both these objects probe physics behind the horizon and grow with time even after thermalization. Both of these proposals have their shortcomings, and many recent studies have tested these proposals in various settings \cite{Susskind3,Susskind4,Susskind5,Susskind6, Barbon, Alishahiha:2015rta, Chemissany,Hol1,Hol2,Hol3,Hol4,Hol5,Hol6, RathU, Hol7, Hol7a, Hol8, Hol9, Hol10, Alish, Hol11,Hol11a, Hol12, Hol13, Hol14, Hol15, Hol16, Hol17, Hol18, Moosa, Hol19, Hol19a, Hol19c,Hol20,Hol21,Hol22,Hol23,Hol24,Hol25,Hol26,Hol19b, Mahapatra, Hol27,Agon, Aliha,Ghodrati,New1}.

Given its importance in holography, it is crucial to be able to quantify complexity in quantum field theory; recently, some progress has been made in this direction \cite{jm, Chapman1, Hashimoto:2017fga,Yang:2017nfn,Khan:2018rzm,Yang:2018nda,Hackl:2018ptj,Alves:2018qfv,Magan:2018nmu,Caputa:2018kdj,Camargo:2018eof, recentmyers, Ame, Yangcomchaos, New, ab}. Computational complexity (or circuit complexity in our case) is an important concept for quantum information theory \cite{c1,c2,c3,c4,CCintroduction,CCintroduction2,Aaronson:2016vto,watrous,osborne2012hamiltonian,gharibian2015quantum}---given a suitable basis, the complexity is the minimum number of operations needed to perform a desired task.
It is of central importance to be able to quantify this, since it helps distinguish the quantum nature of an algorithm from its classical counterpart; this identifies if a proposed quantum computer is indeed a true quantum computer \cite{qsupreme,qsupremea,qsupremeb,qsupremec} \footnote{This list is by no means complete. Interested readers are encouraged to check further references mentioned in these articles and their citations.} . The study of circuit complexity in quantum field theory is in its infancy; only a few cases have been studied to date and much remains unexplored. In \cite{Jordan1,Jordan2,Jordan3}, it was shown that simulation of several field theoretic observables on a quantum computer has an exponential advantage over classical algorithms which use perturbative Feynman diagrams. For our purpose, we will adhere to the notion of complexity associated with a quantum circuit--the task is to prepare the `target state' (for us, this is the time evolved ground state of some Hamiltonian) by a quantum circuit starting from the suitable `reference state', and make this circuit as efficient as possible. 
In \cite{Nielsen1,Nielsen2,Nielsen3}, a geometric approach for circuit complexity was put forward; this was studied in Ref.~\cite{jm} for free scalar field theory. 
Several methods have been proposed/employed to quantify/compute complexity in quantum field theory\cite{jm, Chapman1,Hackl:2018ptj}.
A common feature of these computations is that they all geometrize the quantity. Interestingly, the similarities and differences between these approaches are far from being understood. In this paper, we make some progress in this direction.

Note that one of the key motivations of Susskind et al.~for exploring complexity was that it does not grow quickly with time like the entanglement entropy. The growth sustains over a longer period of time compared to the case of entanglement entropy \cite{Susskind1,Hol8}, and it continues to grow even after the boundary has reached thermal equilibrium until finally saturating at some later time. In the context of holography, the time evolution of complexity has been studied for various types of eternal AdS black holes in different dimensions using both using both the CA and CV conjectures.
Ref.~\cite{Hol12,Hol15} observed that the complexity obtained from CV conjecture monotonically increases with time and then saturates to a (positive) constant, which is reminiscent of the Lloyd bound\cite{c4}. On the other hand, complexity obtained using the CA conjecture for an uncharged black hole remains constant at early times, decreases briefly, and then exhibits a positive growth;
at large times, the complexity saturates, but it does so from above, thus violating the Lloyd bound\cite{Hol12,Hol15}. In Ref.~\cite{Hol5}, a quantity dubbed `complexity of formation` was defined and studied in the context of holography. This quantity is the difference between the complexity associated with the eternal black hole background and the complexity of the vacuum AdS spacetime. In \cite{Hol5}, it was observed that it is divergent for the extremal black hole and a possible interpretation is given based on the fact that states with finite temperature and chemical potential are infinitely more complex than the vacuum state. Additionally, the time evolution of complexity has been studied in the context of collapsing black holes (eternal AdS-Vaidya black hole) using both the CA and CV conjectures \cite{Hol21,Hol24}. Depending on the energy of a collapsing shell, one observes different types of growth patterns for the complexity at early times\cite{Hol24}.  The evolution of complexity has been studied for various other interesting holographic scenarios \cite{Hol1,Hol2,Hol3, Hol8, Hol18, Moosa, Hol14, Hol19a, Hol20, Agon, Mahapatra, Ghodrati, New}. Among these is a proposal to study complexity and its evolution for a subsystem at the boundary \cite{Agon} which extends the notion of complexity for mixed states. A quantity named the `complexity of purification' has been proposed as a diagnostic. While these developments are certainly interesting, they are still at their early stages and a proper field theoretic interpretation is currently outstanding.

A natural place to begin investigating the evolution of complexity is in simple QFTs; some studies have been made in this regard. In \cite{Alves:2018qfv,Camargo:2018eof}, the time evolution of complexity was studied for free scalar field theory after a quench, and a comparison was made with the evolution of the EE. In \cite{New, ab}, this computation has been extended for fermionic systems. In \cite{Camargo:2018eof}, the authors also computed the `complexity of purification' for this model. In \cite{Yang:2018nda,Yangcomchaos}, the nature of complexity evolution from an axiomatic point of view has been discussed. In \cite{Hol15,Yang:2017nfn}, complexity evolution for thermofield double states in CFTs has been studied.  In this work, we consider simple models (free theories) in the hope that this will pave the way forward to study interacting QFTs (appropriate for understanding holography). Second, we would like to differentiate between the various methods for computing complexity in this setup. 
Although our proposal is general, for explicit illustration we will use a generic bosonic lattice model which corresponds to a plethora of interesting QFTs in the continuum limit. We compare the three most common methods of computing complexity, namely (1) the Fubini-Study approach \cite{Chapman1}, (2) the covariance matrix method\cite{Hackl:2018ptj} and (3) circuit complexity (going directly at the wavefunction)\cite{jm}.  We show that only circuit complexity is sensitive to our diagnostic. We further discover/demonstrate a generic pattern of this sensitivity, which hints at interesting physics that might be useful for quantum computation. In \cite{Cap1, Cap2, bartek, cap4,Cap3}, an alternative method of computing complexity using path integral approach has been proposed; in \cite{Taka} its implications for holography were further motivated. We will not consider this path integral approach, but will scrutinize the other three methods discussed above.

To establish our testing method, we will use two common information theoretic measures--the {\sl Loschmidt echo} and {\sl fidelity}. Generally speaking, the Loschmidt echo is defined as the overlap between a reference state and a forward and then backward evolved state. One starts with a reference state $|\psi_0\rangle$, which is first forward evolved by some Hamiltonian followed by a backward evolution by a slightly different Hamiltonian ($|\psi_2\rangle= \exp(i H'_1 t)  \exp(-i H_1 t)  |\psi_0\rangle$). The Loschmidt echo is defined 
as \cite{Echo}
\be \label{LS}
\mathcal{F}_{\text{LE}}= | \langle \psi_0 | \psi_2\rangle | \ .
\ee
This overlap can be thought of as a distance (in state space) between two states. 
Another way to represent the above overlap is the following
\be \label{F}
 \tilde {\mathcal{F}}=|\langle\tilde \psi_1|\psi_1\rangle|,
\ee
where, $|\psi_1\rangle=\exp(-i H_1 t) | \psi_0\rangle$ and $|\tilde \psi_1\rangle=\exp(-i H'_ 1)  |\psi_0\rangle$.

Clearly, these two quantities have the same value, thereby making it insensitive to the details of the evolution of states -- they only depend on the the Hamiltonians $H_1$ and $H'_1$ and the reference state $|\psi_0\rangle$ \cite{Hol8}. These overlaps contain important physical information about the underlying system. In this paper we investigate if there is any difference between ${\cal F}_{\rm LE}$ and $\tilde {\mathcal{F}}$. More explicitly, we address the question--is there any alternative notion of distance that can differentiate between the states involved in these overlaps? Complexity can be a natural candidate for this. To incorporate complexity with this quest we develop a test and then check the different methods of complexities. One of our main results is that circuit complexity from Nielsen method can give us the desired quantity. 
Then we generalize our result for states with multiple evolutions and find a generic property that might be useful from the perspective of quantum computation -- one can compute one quantity with less complexity over the other; this is the quantity which can be simulated more efficiently by a quantum computer.

To quantify the complexity associated with the states involved in ${\cal F}_{\rm LE}$ and $\tilde {\mathcal{F}}$, we will use the `bra' and `ket' of the overlap as the reference and target states. Then we compute the complexities associated with these states for both ${\cal F}_{\rm LE}$ and $\tilde {\mathcal{F}}$ by three methods discussed above. The strength of this method is that we do not need to do a direct comparison between the approaches. Rather, we are exploring the evolution of complexities associates with the states in ${\cal F}_{\rm LE}$ and $\tilde {\mathcal{F}}$ and checking if the evolutions are identical or not.

Computing complexity using the Fubini-Study approach amounts to first identify the target state as some kind of coherent state, and then finding the geodesic (connecting the target and the reference state) distance on this manifold induced by this family of states. On the other hand, both the circuit complexity and the covariance matrix method employ the geometric method  pioneered by Nielsen \cite{Nielsen1,Nielsen2,Nielsen3} and translated in the context of QFT by \cite{jm}, with only difference being, for the first case one uses directly the wave function and the for the second case one uses covariance matrix (appropriate only for Gaussian states). We will discuss them in detail in the later sections.  
We find that only the circuit complexity gives us the desired difference between the complexities defined for the states coming from the Loschmidt echo with those of fidelity; in that sense, circuit complexity is a better approach. Moreover, complexity from the Loschmidt echo is always larger. We extend this idea for an arbitrary number of evolutions and show that the number of evolutions performed on {\it one} state (ket) dictates the complexity between a pair of states, and the state with the highest number of evolutions will have the highest complexity. This implies that if one is interested in overlap measurement between two states, Fidelity which corresponds to the smallest number of evolutions on one state will always be the easiest choice since it has the least complexity. The other two methods are unable to distinguish between the complexities of these two different quantities, thereby demonstrating the advantage of circuit complexity approach over these two methods.

The organization of the paper is as follows. In Section (2) we discuss our model and set up the quench protocol.  In Section (3) we discuss the computation of the complexity for the time evolved ground state of our model by three different methods. In Section (4) we explain our testing procedure and apply this to different methods of complexities.
In the following section we then generalize our arguments and discuss the implications in detail. In Section (6) we briefly explore the time evolution of complexity for non-local theories and compare with results from local theories. Lastly, we conclude by summarizing our results and note interesting future work.
\section{The Model and Quench Protocol}
We consider a free bosonic field theory (regularized) on a lattice;\footnote{We set the lattice spacing to unity.}
the Hamiltonian is
\begin{equation}
 H (q, \hat q)  = \frac{1}{2} \sum_l \left[ p_l^2
     +  q^2\, x_l^2
     + \hat q~ x_{l+1} x_l  \right] \ .
\label{hamilton}
\end{equation}
In Eq.~(\ref{hamilton}), $x_l$ ($p_l$) is the position (momentum) operator at site-$l$, and $\{q, \hat q\}$ parameterize the ``restoring forces"---$q > 0$, but we allow $\hat q $ to have either sign.
This Hamiltonian is more general than a free scalar field theory discretized on a lattice; depending on the choice of parameters,
a variety of interesting behaviors arise.\footnote{E.g. writing $q^2= (a^2+b^2)$ and $\hat q= 2 a\, b$ ($a,b \in {\bf R}$), 
one has the bosonic analog of the Su-Schreiffer-Heeger model \cite{topo1, topomechanics}.}
For us, this provides a convenient/natural medium to explore our testing procedure.



Eq.~(\ref{hamilton}) is readily analyzed by expanding the position and momentum operators in Fourier modes (normal modes) as
\begin{equation} \label{normal}
 x_l = \frac{1}{\sqrt{N}} \sum_k e^{-i\, \frac{2\pi k \, l}{N}} \tilde x_k
 \ \ , \ \ 
 p_l = \frac{1}{\sqrt{N}} \sum_k e^{-i\, \frac{2\pi k \, l}{N}} \tilde p_k
 \ ,
\end{equation}
where $0\leq k \leq (N-1)$ with $N$ being the total number of (lattice) sites;
one obtains\footnote{We have used the orthogonality condition $$\frac{1}{N}\sum_{l=0}^{N-1}\exp[- i\,2\pi\,(k-k')l/N]=\delta_{k,k'}.$$}
\begin{equation}
 H(q,\hat q) = \frac{1}{2} \sum_k \left[ \tilde  p_{k}\tilde p_{-k} 
    + \omega_k^2~ \tilde x_{k}\tilde x_{-k}\right] ,
\label{HBfourier}
\end{equation}
where $\omega_k^2 = q^2  +  \hat q \cos (\frac{2\,\pi\, k}{N})$ and $\omega_k=\omega_{-k}.$ 
Eq.~(\ref{HBfourier}) is then diagonalized by introducing creation and annihilation operators\footnote{$[a^{\phantom \dagger}_k, a_k^{\dagger}]=1$ $\forall$ $k$.}

\begin{equation}
 H(q,\hat q) = \sum_k \omega_k \left( 
  a^{\dagger}_k a^{\phantom \dagger}_k + 1/2 \right) \ ,
\label{Hdiagonal}
\end{equation}
where
\begin{equation}
 \tilde  x_k = \frac{1}{\sqrt{2\omega_k} } \left( 
   a^{\phantom \dagger}_k + a^{\dagger}_{-k} \right)
 \ \ , \ \ 
 \tilde p_k = \frac{1}{i} \sqrt{\frac{\omega_k}{2}}
  \left( a^{\phantom \dagger}_k - a^{\dagger}_{-k} 
  \right).
\label{creationannihilation}
\end{equation}


We are interested in studying quenches in the above model---the quench protocol we employ is 
\begin{subequations}
\begin{eqnarray}
 H & = & H(q,\hat q) \ \ \ \ \ {\rm for} \ t \leq 0  \label{H}
  \\ 
 H & = & H_1(q_1,\hat q_1) \ \ \  {\rm for} \ t > 0 \label{H1} \ ,
\end{eqnarray}
\end{subequations}
where $(q,\hat q)$ and $ (q_1,\hat q_1)$ are different. 
For $t \leq 0$, we prepare the system in the ground state of $H(q,\hat q)$; then we evolve the state by $U_1(t)=\exp[-i\, H_1(q_1,\hat q_1)\, t]$. 
In what follows, we consider the evolution of the complexity following the quench---we consider the complexity between the initial state and the 
time evolved state. 
In the following section, we compute the complexity for this model by the different approaches.


\section{Complexity}

In this section, we will explore the different approaches of probing the complexity. At the end of this section we will comment on the differences between the different approaches. We will investigate the following methods-
\begin{itemize}
\item Complexity from Fubini-Study 
\item Circuit complexity from wave function
\item Circuit complexity from the covariance matrix
\end{itemize}

\noindent
The basic idea of complexity is the following: One starts with a suitable reference state,\footnote{The reference state is preferably an unentangled state in some suitable basis.  We explain in detail the choice of reference state when we discuss the notion of `circuit complexity.'} which one acts on with a set of unitary operators to reach a target state; the complexity corresponds to the minimum number of operations needed to accomplish this.
To carry out this procedure, one first fixes the set of elementary unitary operators and determines the operator space; the complexity is the shortest distance in that (operator) space connecting the reference and the target states \cite{Nielsen1,Nielsen2,Nielsen3}. 
In general, this is a nontrivial and even ambiguous procedure.  To proceed, one defines suitable measures on the operator space, which satisfy the following criteria \cite{jm,Nielsen1,Nielsen2, Nielsen3} : 
\begin{enumerate} 
\item They should be continuous.
\item These should be positive definite. 
\item They should be homogeneous.
\item They should satisfy the triangle inequality. 
\item They can be infinitely differentiable.
\end{enumerate}
Criteria (1)-(4) identify these quantities as legitimate functions to measure distance between points in the underlying space; if condition (5) is satisfied, then they correspond to the distance between points on a Finsler manifold. 
\begin{figure}[ht] 
\centering
 \scalebox{0.40}{\includegraphics{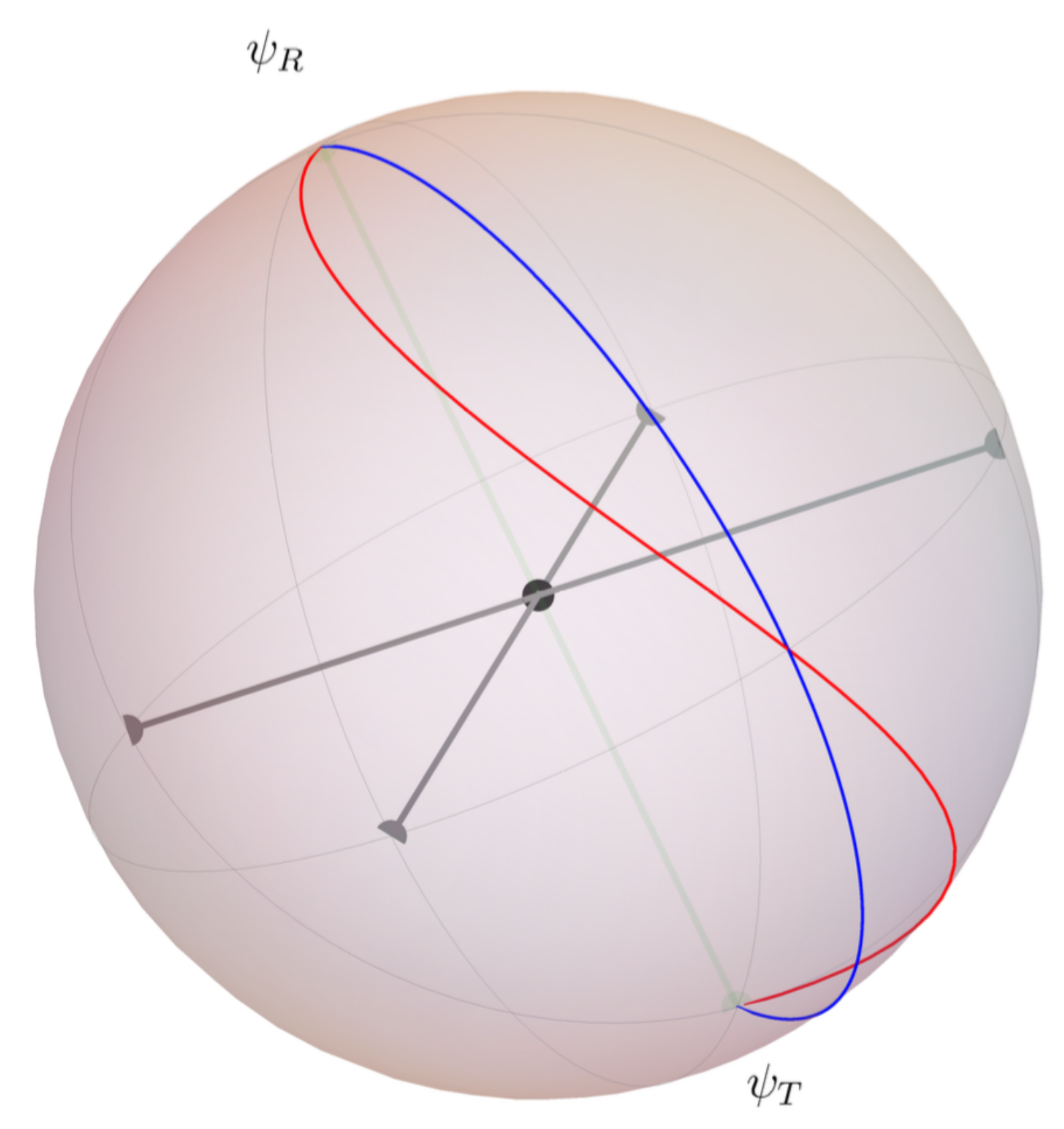} }
\caption{An illustration for different distances between reference ($\psi_R$) and target state ($\psi_T$) in the Hilbert-space. Complexity  (from Fubini-Study approach) will correspond to the length of the geodesic shown in blue.}
\label{fig:1}
\end{figure}

Now there are various methods for computing complexity. The Fubini-Study approach \cite{Chapman1} naturally selects one particular measure. In this method, one typically identifies the state as some kind of coherent state of a particular group, and then one defines a metric for that group manifold; the complexity is computed as the geodesic distance between the reference and target states (for that particular metric). This is illustrated in the Fig.~(\ref{fig:1}), while the details are discussed in Section  (3.1). On the other hand, the `circuit complexity' approach allows one to choose various measures \cite{jm, Hackl:2018ptj,recentmyers} satisfying the properties mentioned above. For the model (and states) considered in this work, it is natural to write the reference and target states in the position representation:
$\psi(\vec v)=\exp(- \vec v^T. A . \vec v)$, where $\vec v=\{x_1,x_2,\cdots\}$ and $A$ is a matrix; then the problem is reduced to finding the optimal unitary which takes the reference 
state ($A^{R}$) to the the target state ($A^{T}$).
\be
A^{T}= U. A^{R}.U^{T}  \ .
\ee
This is discussed in detail in Section (3.2). Now for Gaussian states, they are equivalently described by their covariance matrix; complexity in this case quantifies the minimum number of unitary operations required to generate the covariance matrix of the target state starting from the covariance matrix of the reference state. The details of the covariance matrix approach are presented in Section (3.3).


\subsection{Complexity from Fubini-Study}


In this section, we detail the calculation of the complexity using the Fubini-Study approach.  This is executed by writing the eigenoperators of $H_1$ (the Hamiltonian for $t>0$) in terms of the eigenoperators of $H$ (the Hamiltonian for $t \leq 0$).
As per Eq.~(\ref{creationannihilation}), we have\footnote{Note that $\omega_{1,k}$ and $\omega_k$ are functions of $(q_1,\hat q_1) $ and $(q,\hat q)$, respectively.}
\begin{align}
\begin{split}
&\textrm{for}\,\,  H(q,\hat q) :\,\,\,\, \left(
\begin{array}{c}
 \tilde x_k  \\
 \tilde p_k  \\
\end{array}
\right) =\frac{1}{{\sqrt{2 \omega _k}}
}\left(
\begin{array}{cc}
 1 & 1 \\
 -i\, \omega _k & i\, \omega _k \\
\end{array}
\right)  \left(
\begin{array}{c}
 a_k  \\
 a^{\dagger}_{-k}  \\
\end{array}
\right),\\&
\textrm{for}\,\,  H_1(q_1,\hat q_1) :\,\,\,\, \left(
\begin{array}{c}
\tilde  x_k  \\
\tilde p_k  \\
\end{array}
\right) =\frac{1}{{\sqrt{2 \omega _{1,k}}}
}\left(
\begin{array}{cc}
 1 & 1 \\
 -i\, \omega _{1,k} & i\, \omega _{1,k} \\
\end{array}
\right)  \left(
\begin{array}{c}
 a_{1,k}  \\
 a^{\dagger}_{1,-k}  \\
\end{array}
\right) \ ;
\end{split}
\end{align}
from this, one obtains the Bogoliubov transformation relating $(a_{1,k}, a^{\dagger}_{1,-k})$ to  $(a_{k}, a^{\dagger}_{-k})$:
\begin{align}
\begin{split}
\left(
\begin{array}{c}
 a_{1,k}  \\
 a^{\dagger}_{1,-k}  \\
\end{array}
\right)=\left(
\begin{array}{cc}
 \mathcal{U}_k & \mathcal{V}_k \\
\mathcal{V}_k &  \mathcal{U}_k \\
\end{array}
\right) \left(
\begin{array}{c}
 a_{k}  \\
 a^{\dagger}_{-k}  \\
\end{array}
\right),
\end{split}
\end{align}
where
\begin{equation}
\mathcal{U}_{k}=\frac{\omega_{1,k}+\omega_k}{2\sqrt{\omega_{1,k}\omega_{k}}}
\ \ , \ \ 
\mathcal{V}_{k}=\frac{\omega_{1,k}-\omega_k}{2\sqrt{\omega_{1,k}\omega_{k}}}  \ ,
\nonumber
\end{equation}
with $|\mathcal{U}_k|^2-|\mathcal{V}_k|^2=1$.
Hence, we obtain
\begin{align}
\begin{split} \label{Ham}
H_1(q_1,\hat q_1)=\sum_{k=0}^{N-1}\omega_{1,k}[(\mathcal {U}_{k}^2+\mathcal{V}_{k}^2)\tau_{k}^{z}
 + \mathcal{U}_{k}\mathcal{V}_{k} \tau_{k}^{+} + \mathcal{U}_{k}\mathcal{V}_{k}\tau_{k}^{-}],
\end{split}
\end{align} 
where
\begin{equation}
 \{ \tau_{k}^{+} = a_{k}^{\dagger}a_{-k}^{\dagger} \ \ , \ \ 
     \tau_{k}^{-} = a^{\phantom \dagger}_{-k} a^{\phantom \dagger}_{k} \ \ ,  \ \
     \tau_{k}^{z} = (a_k^{\dagger}a^{\phantom \dagger}_k + a^{\phantom \dagger}_{-k}a_{-k}^{\dagger} )/2  \}
\end{equation}
satisfy an $SU(1,1)$ algebra 
\be
[\tau_{k}^{z}, \tau_{k}^{\pm}]=\pm \tau_{k}^{\pm }
\ \ , \ \ 
[\tau_{k}^{+},\tau_{k}^{-}]=-2\tau_{k}^{z} \ .
\ee

As discussed above, we take the ground state of $ H(q,\hat q,q') $ as our reference state; this is given by
\be \label{ground}
|\psi_0\rangle=\prod_{k=0}^{N-1}|k,-k\rangle \ ,
\ee
where $|k,-k\rangle$ denotes the Fock vacuum for modes $k$ and $(-k)$.  
We are interested in the complexity of the time-evolved state
\be \label{evolv}
|\psi_1(t)\rangle= U_1(t)|\psi_0 (t=0)\rangle.
\ee
To evaluate this, we employ the decomposition\footnote{Now $\tau_k^{-}$ simply annihilates the state $|k, -k\rangle$ for all values of $k.$ So the action of $\exp(\gamma_k^{-}\tau_k^{-})$ on $|k, k\rangle$ is trivial. Also, $\tau_k^{-} |k, -k\rangle= \frac{1}{2} |k,-k\rangle.$ Then, $\exp((\ln \gamma_k^0)\tau_k^{z}) |k,-k\rangle=(\gamma_k^0)^{1/2}|k, -k\rangle.$ This produces  just an overall phase and can be absorbed inside the normalization ($\mathcal{N}_k(t)$) of the state. Non trivial effects comes from the $\exp(\gamma_k^{+}\tau_k^{+})$.}\cite{perelomov}
 \begin{align}
 \begin{split} \label{BCH}
\exp(\beta_k \, \tau_k^z+\alpha_k^{+}\tau_k^{+}+\alpha_k^{-}\tau_k^{-})=\exp(\gamma_k^{+}\tau_k^{+})\exp ((\ln \gamma_k^0)\tau_k^{z})\exp(\gamma_k^{-}\tau_k^{-}),
  \end{split}
 \end{align}
 where 
\begin{equation} \label{def1}
 \gamma_k^0=\Big(\cosh(\mu_k)-\frac{\beta_k}{2\mu_k}\sinh(\mu_k)\Big)^{-2}
 \ \ , \ \ 
 \gamma_k^{\pm}=\Big(\frac{\alpha_k^{\pm}}{\mu_k}\Big)\Big ( \frac{\sinh(\mu_k)}{\cosh(\mu_k)-\frac{\beta_k}{2\,\mu_k}\sinh(\mu_k)} \Big )
 \end{equation}
with $\mu_k^2=(\beta_k^2/4)-\alpha_k^{+}\alpha_k^{-}$; 
one obtains 
\begin{equation} \label{evolv1}
 |\psi_1(t)\rangle = 
     \prod_{k=0}^{N-1}\mathcal{N}_{k}(t)\exp(\gamma_{1, k}^{+}(t)\, a^{\dagger}_{k}a^{\dagger}_{-k})|k,-k\rangle \ ,
 \end{equation}
where
\begin{equation}  \label{def}
\gamma_{1, k}^{+}= \Big(\frac{\alpha_{1,k}^{+}}{ \mu_{1,k}}\Big) 
       \Big ( \frac{\sinh(\mu_{1,k})}{\cosh(\mu_{1,k})-\frac{\beta_{1,k}}{2\,\mu_{1,k}}\sinh(\mu_{1,k})} \Big )
  \ \ , \ \ 
\mu_{1,k}^2=\frac{\beta_{1,k}^2}{4}-\alpha_{1,k}^{+}\alpha_{1,k}^{-}  \ , \\
\end{equation}
with
\begin{equation}
\beta_{1,k}=-i\, t\, \omega_{1,k}\, (\mathcal{U}_k^2+\mathcal{V}_k^2)
  \ \ , \ \ 
\alpha_{1,k}^{+}=\alpha_{1,k}^{-}=-i\,t\,\omega_{1,k} \, \mathcal{U}_k\mathcal{V}_k \ .
\nonumber
\end{equation}
The state Eq.~(\ref{evolv1}) can be thought of as an $SU(1,1)$ coherent state; the state manifold can be given a Riemannian structure\cite{bookt} -- 
considering the class of states
\be \label{class} 
 |\psi\{\gamma_{k,\tau}(t)\}\rangle
 =\prod_{k=0}^{N-1}\mathcal{N}_k(t)\exp(\gamma_{k,\tau}(t) a^{\dagger}_k\, a^{\dagger}_{-k})|k,-k\rangle 
\ee 
labeled by the parameter $\tau$ and evaluating the Fubini-Study line element
\begin{equation}
 \left( \frac{ds}{d\tau} \right)^2 = \langle \frac{d\psi}{d\tau} \mid \frac{d\psi}{d\tau} \rangle
  - \langle \frac{d\psi}{d\tau} \mid \psi \rangle \langle \psi \mid \frac{d\psi}{d\tau} \rangle  \ ,
\label{FS}
\end{equation}
one obtains
\be \label{met}
ds^2=\sum_{k=0}^{N-1} ds_k^2=\sum_{k=0}^{N-1}\frac{|d\gamma_{k,\tau}|^2}{(1-|\gamma_{k,\tau}|^2)^2}.
\ee
For each value of $k$, one has $H^2$ in the $CP^1$ representation.

For a given $k$, the distance is naturally defined by\footnote{For simplicity, we defined range of $\tau$ from $0$ to $1$; one can reparametrize it to redefine its initial and final value.} 
\be
s_k=\int_{0}^{1} d\tau\, \frac{1}{1-|\gamma_{k,\tau}|^2}\Big |\frac{d\gamma_{k,\tau}}{d\tau}\Big|.
\ee
The full state manifold has the form $H^2\times H^2\times \cdots$ -- the distance can be defined as
\be
s=\sqrt{\sum_{k=0}^{N-1} s_k^2} \ .
\label{distancefunction}
\ee
In this (Fubini-Study) approach, the complexity geodesic distance between the reference and target states (\ref{com1}) --
it follows from Eq.~(\ref{distancefunction}) that the complexity is
\be \label{com1}
\mathcal{C}_{FS}= \sqrt{\sum_{k=0}^{N-1}\mathcal{C}_{k}^2} \ ,
\ee
where $\mathcal{C}_k$ is the geodesic distance for a particular $k$.

To proceed, we write
\be
\gamma_{k,\tau}=|\gamma_k| \exp(i\,\phi_k)
\ \ , \ \ 
|\gamma_k|=\tanh\Big(\frac{\theta_k}{2}\Big) \ ;
\ee
one obtains
\be
ds^2=\frac{1}{4}\,\sum_{k=0}^{N-1} (d\theta_k^2+\sinh(\theta_k)^2 d\phi_k^2).
\ee
Considering two points $(\theta_{1,k},\phi_{1,k})$ and $(\theta_{2,k},\phi_{2,k})$, the complexity (\ref{com1}) takes the form
\be \label{genFS}
 \mathcal{C}_{FS} = \frac{1}{2}\sqrt{ \sum_{k=0}^{N-1}\Big(
   \arcosh\Big[\cosh(\theta_{1,k})\cosh(\theta_{2,k}) - \sinh(\theta_{1,k})\sinh(\theta_{2,k})\cos(\phi_{1,k}-\phi_{2,k})\Big] \Big)^2} \ .
 \ee
For reference and target states given by Eqs.~(\ref{ground}) and (\ref{evolv1}), respectively, $\theta_{1,k}=0$ and $\theta_{2,k}=2\arctanh |\gamma_{1,k}|$ with 
$\gamma_{1,k}$ defined in (\ref{def}),\footnote{For a more detailed discussion about the choice of reference state, see Appendix~(B).}
the complexity takes the form
 \be \label{circfs}
 \mathcal{C}_{FS}=\sqrt{\sum_{k=0}^{N-1} (\arctanh |\gamma_{1,k}|)^2}.
 \ee

\subsection {Circuit Complexity \'a la Nielsen}

We now detail the calculation of the circuit complexity \footnote{In the rest of the text whenever we  will mention circuit complexity we will refer to  this approach.}.  This approach was pioneered by Nielsen;\cite{Nielsen1} it was adapted for free scalar field theory in \cite{jm} and has recently been generalized for interacting field theories in \cite{Ame}.
We  start with the (defining) expression
\begin{align}
\begin{split}
|\psi^{T}_{\tau=1}\rangle=\tilde U(\tau)|\psi^{R}_{\tau=0}\rangle,
\end{split}
\end{align}
where $U(\tau)$ is a unitary operator representing the quantum circuit, which takes the reference state $|\psi^{R}\rangle$ defined 
at $\tau=0$ to the target state $|\psi^{T}\rangle$ defined at $\tau=1$.
As before, $\tau$ parametrizes a path in the Hilbert space (and one can re-parametrize this $\tau$ in various ways). 
Now the unitary operator can be written as a path-ordered exponential
\be \label{Unit}
\tilde U(\tau)= {\overleftarrow{\mathcal{P}}}\exp(i\int_{0}^{\tau }\, d\tau\, H(\tau)),
\ee
where $H(\tau)$ is a Hermitian operator. Next we fix a basis $\{ M_{I} \}$ and expand $H(\tau)$ in term of this basis:
\begin{equation}
 H(\tau)= Y^{I}(\tau) M_{I} \ ,
\nonumber
\end{equation}
where the coefficients $\{ Y^{I}(\tau) \}$ are referred to as `control functions'.  
The $\{ M_I \}$ provide the elementary gates that will be used. The algebra satisfied by these gates gives us the structure of the group; the unitary $\tilde U(\tau)$ can be parametrized as a general element of that group. The goal now is to minimize the depth of this circuit to find the optimal control functions $\{ Y^{I}(\tau) \}$.  To this end, we define the circuit complexity 
$\mathcal{C}(\tilde U)$ through suitable \textit{cost functions} $\mathcal{F}(\tilde U, \dot {\tilde U})$ as\cite{jm,Nielsen1,Nielsen2,Nielsen3}
\be \label{cost}
\mathcal{C}(\tilde U)=\int_{0}^{1} \mathcal{F}(\tilde U,\dot{\tilde U})\, d\tau \ .
\ee
We minimize this cost function and find the geodesic connecting the two states; evaluating $\mathcal{C}(\tilde U)$ on this geodesic, we obtain the complexity. There are various possible choices for these  functions $\mathcal{F}(\tilde U,\dot{\tilde U})$, but they should satisfy the conditions (1)-(5) discussed in Section (3). Here we mention a few of them that have been used extensively in the literature:\cite{jm, Hackl:2018ptj,recentmyers}
\begin{align}
\begin{split} \label{meas}
& \mathcal{F}_2(U,Y)=\sqrt{\sum_{I}p_{I} (Y^{I})^2}, \mathcal{F}_{\kappa}(U,Y)=\sum_{I}p_{I}\, |Y^{I}|^{\kappa}, \quad \kappa\textrm{\text{\ is\, an\, integer \&}} \ \kappa \geq 1,\\& \mathcal{F}_{p}(U,Y)=(\tr ( V^{\dagger} V)^{p/2} ))^{1/p},  V= Y^{I}(\tau) M_{I},  \quad p \textrm{\quad is\, an\, integer}\,\footnotemark.
\end{split}
\end{align}
\footnotetext{These are formally known as `Schatten Norms' and first considered in \cite{Hackl:2018ptj}  and explored in detail in \cite{recentmyers}.}
\noindent
The $\{p_{I} \}$, known as `penalty factors', are weights which, at the moment, are arbitrary.
Among these, $\mathcal{F}_{\kappa=1}$ directly counts the number of gates; most importantly, $\mathcal{F}_{2}$ with $p_I=1$ for all $I$ is basically a distance function on a given manifold. 
We note that the complexity computed using $\mathcal{F}_{2}$ is very similar to $\mathcal{C}_{FS}$, as both of are coming from evaluating the shortest between two points on a given manifold; the difference lies in the fact that circuit complexity, a priori, cannot fix the $\{ p_I \}$ -- we have to make a choice for that.  The Fubini-Study approach canonically fixes it (in fact, they are all fixed to unity)\cite{Chapman1}. 
In the subsequent analysis, we compute the complexity using $\mathcal{F}_2$ (with $p_I=1$ for all $I$) to make a direct comparison with the Fubini-Study approach.

For our case the target wave function following from (\ref{evolv}) is given by\footnote{See Appendix (A) for further details.}
\begin{align}
\begin{split} \label{tarw}
\psi^{\tau=1} (\tilde x_k,t) =\mathcal{N}^{\tau =1} (t) \, \exp \left [- \frac{\sum_{k=0}^{N-1} \Omega_{k} \, \tilde x_{k}^2}{2} \right] \ ,
\end{split}
\end{align}
where $\mathcal{N}^{\tau =1} (t)$ is the normalization factor. 
The frequencies ($\Omega_k$) are given by
\be \label{omegak}
\Omega_{k}=\omega_{1,k} \Big [\frac{\omega_{1,k}-i\,\omega_k \cot (\omega_{1,k}\, t)}{\omega_k-i\,\omega_{1,k}\cot (\omega_{1,k}\, t)}\Big] \ ;
\ee
their real and imaginary parts are
\begin{align}
\begin{split} \label{smallt}
&\Re(\Omega_k)=\frac{\omega_{1,k}^2\omega_k}{\sin(\omega_{1,k}\,t)^2(\omega_k^2+\omega_{1,k}^2\cot(\omega_{1,k}\,t)^2)},\\&
\text{Im} (\Omega_k)=\frac{\omega_{1,k}(\omega_{1,k}^2-\omega_k^2)\sin(2\, \omega_{1,k}\,t)}{2\sin(\omega_{1,k}\,t)^2(\omega_k^2+\omega_{1,k}^2\cot(\omega_{1,k}\,t)^2)}.
\end{split}
\end{align}
This wave function can be written as 
\be \label{tarw1}
\psi^{\tau=1} (\tilde x,t) =\mathcal{N}^{\tau=1}(t)  \, \exp \left [-\frac{1}{2}\Big(v_a. A^{\tau=1}_{ab}. v_b\Big)\right] \ , 
\ee
where $v=\{\tilde x_0, \cdots \tilde x_{N-1}\} $ and $A^{\tau=1}$ is an $ N \times N$ diagonal matrix
\be
A^{\tau=1}=\text {diag} \{\Omega_{0},\cdots,\Omega_{k}\}.
\ee
 Also we take the reference as  (in the same basis $v$)
\be \label{refw}
A^{\tau=0}=\text{diag} \{\omega_{r},\cdots,\omega_{r}\}.
\ee
$\omega_{r}$ can in general be complex. The unitary (\ref{Unit}) acts as
\be \label{tarw2}
A^{\tau}= \tilde U(\tau). A^{\tau=0}. \tilde U^{T}(\tau)
\ee
with the boundary conditions
\be \label{circbound}
A^{\tau=1}= \tilde U(\tau=1). A^{\tau=0}. \tilde U^{T}(\tau=1)
\ \ , \ 
A^{\tau=0}= \tilde U(\tau=0). A^{\tau=0}. \tilde U^{T}(\tau=0) \ . 
\ee

A convenient way to parametrize this $\tilde U(\tau)$ is as follows ,
\be
\tilde U(\tau)=\overleftarrow{\mathcal{P}}\exp\Big(\int_{0}^{\tau} Y^{I}(\tau)M_{I} d\tau\Big) \ ,
\ee
where at $\tau=1$ we reach the target state.  Now the components of  $A^{\tau=0}$ and $A^{\tau=1}$ for our case can be complex. So we restrict ourselves to $GL(N,C)$ unitary. Then  control functions  $\{ Y^{I} \}$ are complex parameters and the $\{ M^{I} \}$ are the $N^2$ generators. Then,
\be \label{tarw3}
Y^{I}=\tr(\partial_{\tau}\tilde U(\tau). \tilde U(\tau)^{-1}. (M^{I})^{T}) \ ,
\ee
where $\tr (M^{I}. (M^{J})^{T})=\delta^{IJ}$ and $ I,J=0,\cdots, N^2-1$. Then the metric can be defined as, 
\be \label{tarw4}
ds^2= G_{IJ}dY^{I} dY^{* J}.
\ee
There is a certain arbitrariness regarding the choice of $G_{IJ}$---we choose, for simplicity, $G_{IJ}=\delta_{IJ}$\cite{jm} i.e. we are fixing the penalty factors to unity; this will enable us to make a more direct comparison with the Fubini-Study approach.

Since we are working with a basis in which both the reference and the target states can be simultaneously diagonalized, the off-diagonal components coming  from some of the elements of  $GL(N,C)$ will increase the distance between states; the shortest distance corresponds to them being set to zero \cite{jm}.  Hence, the $\tilde U(\tau)$ will take the form
\be \label{newnew}
\tilde U(\tau)=\exp\left(\sum_{k=0}^{N-1} \alpha^k(\tau) M_k^{diag}\right) \ ,
\ee
where the $\{ \alpha^{k} (\tau) \}$ are complex, and the $\{ M_k^{diag} \}$ are the ($N$) diagonal generators containing only one identity at the $k$'th diagonal entry. 
Then using (\ref{tarw4}), one obtains the flat metric
 \be \label{circmet}
ds^2=\sum_{k=0}^{N=1}( (d\alpha^{k,1} )^2+(d\alpha^{k,2})^2) \ ,
 \ee
where the superscripts $1$ and $2$ denote the real and imaginary part of $\alpha^k$, respectively.  
It follows the geodesic is simply a straight line of the form
\be \label{Newsol1}
\alpha^{k,j}(\tau)=\alpha^{k,j}(\tau=1)\,\tau+\alpha^{k,j}(\tau=0)
\ee
for each value of $k$ ($j=1,2$); using the boundary conditions, one obtains
\begin{align}
\begin{split} \label{newnew1}
&\alpha^{k,1}(\tau=0)=\alpha^{k,2}(\tau=0)=0,\\&
 \alpha^{k,1}(\tau=1)=\frac{1}{2}\log\frac{|\Omega_k|}{|\omega_r|}
 \ \ , \ \ 
\alpha^{k,2}(\tau=1)=\frac{1}{2}\arctan\frac{\Re(\omega_r)\text{Im}(\Omega_k)-\Re(\Omega_k)\text{Im} (\omega_r)}{\Re(\omega_r)\Re(\Omega_k)+\text{Im} (\Omega_k)\text{Im} (\omega_r)}
\end{split}
\end{align}
for each $k$.
Then the complexity is given by
\be
\mathcal{C}(\tilde U)=\int_{0}^{1}ds\sqrt{ g_{ij}\dot x^{i}\dot x^{j}} \ ,
\ee
where $g_{ij}$ denote the components of the metric (\ref{circmet}), and the $x^{i}$'s are coordinates associated with this metric.
Finally, one obtains
\begin{align}
\begin{split} \label{genCirc}
\mathcal{C}(\tilde U)&=\frac{1}{2}\, \sqrt{
 \sum_{k=0}^{N-1} \left[\Big(\log\frac{|\Omega_k|}{|\omega_r|} \Big)^2+ \Big(\arctan\frac{\Re(\omega_r)\text{Im} (\Omega_k)-\Re(\Omega_k)\text{Im} (\omega_r)}{\Re(\omega_r)\Re(\Omega_k)+\text{Im} (\Omega_k)\text{Im} (\omega_r)}\Big)^2 \right ]}.
\end{split}
\end{align}

Like before (i.e in Section (3.1)) we choose the reference as the ground state of $H(q,q')$ at $t=0$ --- $\omega_r$ will be $\omega_k$ as defined in (\ref{Hdiagonal}); we obtain
\be \label{circ}
\mathcal{C}(\tilde U)=\frac{1}{2}\, \sqrt{
 \sum_{k=0}^{N-1} \left[\Big( \log \frac{ |\Omega_k | }{\omega_{k} } \Big)^2+ \Big( \arctan \frac{\text{Im} (\,\Omega_k )}{\Re( \Omega_k)}\Big)^2 \right ]} \ .
\ee
This expression is very pleasing---the first part is the logarithm of the ratio of the frequencies of target and reference state,\footnote{Again, for a more detailed discussion about the choice of reference state, see Appendix (B).} and is similar to time independent case \cite{jm}; however, we have an additional contribution from the phase term, namely the second term in (\ref{circ}). This is very reasonable, since the time-evolved state has a non-trivial phase. To reproduce those phases starting form a simple reference state, one needs appropriate unitary operators and they will obviously generate certain cost; the complexity evaluated by this method is aptly capturing that.

Next we look at the structure of the optimized circuit. We rewrite (\ref{newnew}) using (\ref{newnew1}) and get,
\be 
\tilde U(\tau)=\prod_{k=0}^{N-1} \exp\Big[(\tilde \alpha^{k}M^{diag}_{k} +\tilde\beta^{k}\tilde M^{diag}_{k})\,\tau \Big] .
\ee
\noindent
Here $k$ runs from  $0$ to $N-1$ and the corresponding $\tilde \alpha^{k}$'s are equal to $\frac{1}{2} \log \frac{ |\Omega_k | }{\omega_{k}}$. 
$\tilde M^{diag}_{k}$ are $-2\,i$ times the diagonal generators and $\tilde \beta^{k}$'s are $-\frac{1}{4}\arctan \frac{\text{Im} (\,\Omega_k )}{\Re( \Omega_k)}$ for corresponding values of $k.$ Now we can identify the unitary gates which are the building blocks of $\tilde U(\tau).$ The unitary operators take the form $\exp(i\, O),$ where $O$ is some Hermitian operator. Now given the basis $v=\{\tilde x_0,\cdots, \tilde x_{N-1}\}$, we can act on it by the unitary operators and from their action on $v$ we can identify $(i\,O)$ to these matrices $M_k^{diag}$ and $\tilde M_k^{diag}.$ The first one $M_{k}^{diag}$ corresponds to $(i\, \tilde x_{k}\tilde p_{k})$ for each value of $k.$ They are the usual scaling operators and there are $N$
of them. The  second one $\tilde M_{k}^{diag}$ corresponds to $\{i\, (\tilde x_{k}\tilde p_{k})(\tilde x_k\,\tilde p_k)+\tilde x_k\,\tilde p_k\}$. In terms of these operators the optimal circuit that generates the required target state takes the following form,
\begin{align}
\begin{split} \label{finunit}
\tilde U(\tau)=\prod_{k=0}^{N-1}\exp (i\,\epsilon\, \tilde \alpha^k\,\tau\, O_k)\exp(i\,\epsilon\,  \tilde \beta^k\,\tau\,\hat O_k),
\end{split}
\end{align}
where \be \label{difop} O_{k}=(\tilde x_k\,\tilde p_k+\tilde p_k\,\tilde x_k), \hat O_k= O_k\, O_k\ee and $\tilde \alpha^{k}=\frac{1}{2\,\epsilon} \log \frac{ |\Omega_k | }{\omega_{k}}, \tilde \beta^{k}=-\frac{1}{4\,\epsilon}\arctan \frac{\text{Im} (\,\Omega_k )}{\Re( \Omega_k)}$. Note that here we have introduced an infinitesimal parameter $\epsilon$. For all practical purposes (at least from the point of view of implementation), the target state can only be achieved up to a certain tolerance, i.e  $| |\psi^{T}\rangle-U|\psi^{R}\rangle| < \epsilon$. Basically, $\epsilon$ plays the role of a small error, which we want to minimize as much as possible \cite{jm}. $O_k$s scale the coefficients of $\tilde x_i^2$ inside the wave function by a real number and the other operator $\hat O_{k}$ scales the coefficients of $\tilde x_i^2$ by a complex number. Together they are sufficient to reproduce the target wave function as the coefficients of $\tilde x_i^2$ inside the wave function are complex numbers. \par
Note that, one could have used $i\, \tilde x_{i}^2$  together with the scaling operators to get the target state from the reference state.  But the geodesic analysis prefers rather different set of operators beside the scaling operators. The reason is two fold. As we have seen, when the target $A^{\tau=1}$ and the reference $A^{\tau=0}$ can be simultaneously diagonalized, the geodesic is just a straight line path and that, in turn, gives us the optimal circuit consisting of mutually commuting gates \cite{jm}. From this one can easily see that it rules out the possibility of having $i \tilde x_i^2$ and scaling together, since they don't commute with each other. The second reason is a technical reason. Given the basis $v=\{\tilde x_0,\cdots, \tilde x_{N-1}\}$ we can see that it is not possible to write down a matrix representation for the operator $e^{i\, \tilde x_{i}^2}$ as the action of $i \, \tilde x_i^2$ on the basis $v$ takes it out of the basis. In other words, the action of this operator on the reference state is non-linear.  On the other hand, we will see in the next section that one can find a representation of $i \tilde x_i^2$, with respect to the covariance matrix and the optimal unitary coming from the geodesic analysis will consist of this operator. 


\subsection{Circuit Complexity from the covariance matrix}

Just like the reference wave function, the target wave function (\ref{tarw}) is purely Gaussian; it can be completely characterized by its first and second moments. 
Therefore, we can define a covariance matrix \cite{bookt1}, which will contain the same information as the matrix $A$ defined in the wave function (Eq. (\ref{tarw1})); 
hence, we can reformulate the analysis of Section (3.2) in terms  of the covariance matrix \cite{Hackl:2018ptj}.

The components of the covariance matrix ($G$) can be defined via
\be
G_{ab}=\langle \psi(\tilde x_k,t)|\xi_a\xi_b+\xi_b\xi_a|\psi(\tilde x_k,t)\rangle,
\ee
where
$\vec \xi=\{\tilde x_0, \tilde p_0,\cdots, \tilde x_{N-1},\tilde p_{N-1}\}.$ 
The matrix $G$  then takes the form
\be
G=\left(
\begin{array}{cccc}
 (G^{k=0})_{2\times 2} & \cdots & \,\, 0 \\
 \vdots&   \ddots& \ddots \\
 0 &\cdots & (G^{k=N-1})_{2\times 2} \\
\end{array}
\right).
\ee
For each value of $k$, the matrix $G$ factorizes further into $2\times 2$ symmetric blocks---these blocks have one-to-one corresponds with the canonical pair $\{\tilde x_k, \tilde p_k\}$;
hence, there will be $N$ of the $2\times 2$ blocks. The matrix $G$ is $(N\times N)$ symmetric matrix. For our target state (\ref{tarw}), each of the $(2\times 2)$ matrices takes the form
\be
G^{\tau=1\,, k}_{2\times 2} =\left(
\begin{array}{cc}
 \frac{1}{ \Re  (\Omega_k)}& -\frac{\text{Im} (\Omega_k)}{\Re (\Omega_k)} \\
-\frac{ \text{Im} (\Omega_k)}{\Re  (\Omega_k)} & \frac{|\Omega_k|^2}{\Re (\Omega_k)} \\
\end{array}
\right).
\ee
Also for the reference state (\ref{refw}) we have
\be
G^{\tau=0\,, k}_{2\times 2} =\left(
\begin{array}{cc}
 \frac{1}{ \Re  (\omega_r)}& -\frac{\text{Im} (\omega_r)}{\Re (\omega_r)} \\
-\frac{ \text{Im} (\omega_r)}{\Re  (\omega_r)} & \frac{|\omega_r|^2}{\Re (\omega_r)} \\
\end{array}
\right) \ .
\ee
Note that the determinants of both of these matrices are unity. 
Now to facilitate the computation we will perform a change of basis for each of the smaller blocks as follows.
\be
\tilde G^{\tau=1\,,k}= S. G^{\tau=1\,,k}. S^{T}, \tilde G^{\tau=0\,,k}= S. G^{\tau=0\,,k}. S^{T}, 
\ee
with\footnote{This is just one possibility---we could have chosen differently; that would have also produced the desired result.} 
\be S=\frac{1}{\sqrt{\Re(\omega_r)(\text{Im} (\omega_r)^2+(\Re(\omega_r)-1)^2)}}
\left(\begin{array}{cc}
|\omega_r|^2-\Re(\omega_r) & \text{Im} (\omega_r)\\
\text{Im} (\omega_r)&1-\Re(\omega_r)\\
\end{array}
\right),
\ee
such that $\tilde G^{\tau=0\,,k}= I$ (an identity matrix). Therefore, $\tilde G^{\tau=1\,,k}$ and $\tilde G^{\tau=0\,,k}$ always commute with each other and can be diagonalized simultaneously \cite{Camargo:2018eof}. This is same as Section~(3.2), where $A^{\tau=1}$ and $A^{\tau=0}$ commute with each other. In terms of the covariance matrix the statement (\ref{tarw2}) becomes,
\be 
\tilde G^{\tau}= \tilde U(\tau). \tilde G^{\tau=0}. \tilde U^{T}(\tau).
\ee
Now as before we restrict ourselves to the $GL(R)$ unitary. Given the fact that $G$'s admit a block structure, as before it is
convenient to parametrize $\tilde U(\tau)$ as GL(2N, R)=GL(2, R)$\times \cdots (2N-2)  \cdots \times$GL(2, R)\footnote {This is a special choice, but given the block diagonal structure of $\tilde U(\tau)$ we can easily justify this.}.  Further we can parametrize each of these $GL(2, R)$  block as $ R \times SL(2, R).$ We will now  conduct all the subsequent analysis block by block. For each block we have, 
\begin{align}
\begin{split}
\tilde U^{k}(\tau)=\exp(y_k(\tau)) \, \left(
\begin{array}{cc}
\cos \phi_k(\tau) \cosh \rho_k(\tau)-\sin \theta_k(\tau) \sinh \rho_k(\tau) & -\sin \phi_k(\tau) \cosh \rho_k(\tau) +\cos \theta_k(\tau) \sinh \rho_k(\tau)\\
\sin \phi_k(\tau) \cosh \rho_k(\tau)+\cos \theta_k(\tau) \sinh \rho_k(\tau) &  \cos \phi_k(\tau) \cosh \rho_k(\tau) +\sin \theta_k(\tau) \sinh \rho_k(\tau)\\
\end{array}
\right).
\end{split}
\end{align} 
Next we set the boundary conditions as before. 
 \be
\tilde G^{\tau=1\,,k}=\tilde U^{k}(\tau=1). \tilde G^{\tau=0\,,k} . (\tilde U^{k}(\tau=1))^{T}.
\ee
This will give the final boundary condition as follows
\begin{align}
\begin{split}
&\{y_k(\tau=1),\cosh 2\rho_k(\tau=1), \tan (\theta_k(\tau=1)+\phi_k(\tau=1)) \}\\&=\{0,\frac{\Re(\omega_r)^2+\Re(\Omega_k)^2+(\text{Im} (\omega_r)-\text{Im} (\Omega_k))^2}{2\,\Re(\omega_r)\,\Re(\Omega_k)},\,\frac{\tilde G_{11}^{\tau=1,\,k}-\tilde G_{22}^{\tau=1,\,k}}{2\,\tilde G_{12}^{\tau=1,\,k}}\}, 
\end{split}
\end{align}
where $\tilde G_{ij}^{\tau=1,\,k}$ denote various components of the matrix $\tilde G^{\tau=1,\,k}.$
Also we need the following,
\be
\tilde G^{\tau=0\,,k}=\tilde U^k(\tau=0).\tilde  G^{\tau=0\,,k} . (\tilde U^k(\tau=0))^{T}.
\ee
This gives, 
\be
\{y_{k}(\tau=0),\rho_{k}(\tau=0), \theta_k(\tau=0)+\phi_k(\tau=0)\}=\{0,0, c_k\}.
\ee
$c_k$ is an arbitrary. For simplicity we will choose $\phi_k(\tau=1)=\phi_k(\tau=0)=0$ and $\theta_k(\tau=1)=\theta_k(\tau=0)=c_k=\tan^{-1}\Big(\frac{\tilde G_{11}^{\tau=1,\,k}-\tilde G_{22}^{\tau=1,\,k}}{2\,\tilde G_{12}^{\tau=1,\,k}}\Big).$ Given this form of $\tilde U^k(\tau)$ we can write down the metric as shown in (\ref{tarw3}) and (\ref{tarw4}). Following\cite{jm, Hackl:2018ptj,recentmyers} the metric is (we have set $G_{IJ}=\frac{1}{2}\delta_{IJ}$ for simplicity),
\be \label{met2}
ds_k^2=dy_{k}^2+ d\rho_k^2+\cosh(2\rho_k)\cosh^2\rho_k \,d\phi_k^2+\cosh(2\rho_k)\sinh^2\rho_k\, d\theta_k^2-\sinh(2\rho_k)^2\,d\phi_k d\theta_k.
\ee
The total complexity (after summing over $k$) is defined as,
\be
\mathcal{C}(\tilde U)=\int_{0}^{1}ds\sqrt{\sum_{k=0}^{N-1} g^{k}_{ij}\dot x_k^{i}\dot x_k^{j}},
\ee
where $g_{ij}^{k}$ denote the components of the metric (\ref{met2}) for each $k$ and $x^{i}$'s are coordinates associated with this metric for each value of $k.$
The simplest solution for the geodesic is again a straight line on this geometry\cite{jm}. 
\be \label{solsol}
y_{k}(\tau)=0, \rho_{k}(\tau)=\rho_k(\tau=1)\,\,\tau,\theta_{k}(\tau)=\theta_{k}(\tau=0),\phi_{k}(\tau)=0.
\ee
So finally we get ,
\begin{align}
\begin{split} \label{genCov}
\mathcal{C}(\tilde U)=\sqrt{\sum_{k=0}^{N-1} \rho_k(\tau=1)^2}=\frac{1}{2}\sqrt{\sum_{k=0}^{N-1}\left(\arcosh \left (\frac{\Re(\omega_r)^2+\Re(\Omega_k)^2+(\text{Im} (\omega_r)-\text{Im}(\Omega_k))^2}{2\,\Re(\omega_r) \Re(\Omega_k)}\right)^2\right)}.
\end{split}
\end{align}
Choosing the reference state as the ground state of (\ref{H}), this simplifies to the following,
\begin{align}
\begin{split} \label{comCov}
\mathcal{C}(\tilde U)=\frac{1}{2}\sqrt{\sum_{k=0}^{N-1}\left(\arcosh \left (\frac{\omega_k^2+|\Omega_k|^2}{2\,\omega_k\, \Re(\Omega_k)}\right)^2\right)}. 
\end{split}
\end{align}
Now let's investigate the structure of the optimal circuit. Give this solution (\ref{solsol}) we get,
\begin{align}
\begin{split}
\tilde U^{k}(\tau)=\exp\Big[\tilde M\, \tau\Big],
\end{split}
\end{align}
where
\be
\tilde M=\left(
\begin{array}{cc}
-\sin \theta_k(\tau=0) & \cos \theta_k(\tau=0)\\
\cos \theta_k(\tau=0) &  \sin \theta_k(\tau=0) \\
\end{array}
\right)\rho_{k}(\tau=1).
\ee
This can be decomposed in terms of SL(2, R) generators in the following way,
\be
\tilde M =\alpha_1 M_{11}+ \alpha_2 M_{22}+\alpha_3 M_{33},
\ee
where
\begin{align}
\begin{split}
M_{11}=\left(
\begin{array}{cc}
-1 &0\\
0& 1 \\
\end{array}
\right), M_{22}=\left(
\begin{array}{cc}
0 &0\\
1& 0 \\
\end{array}
\right),M_{33}=\left(
\begin{array}{cc}
0 &1\\
0& 0 \\
\end{array}
\right),\\
\alpha_1=\sin \theta_k(\tau=0),\alpha_2=\alpha_3=\cos \theta_k(\tau=0).
\end{split}
\end{align}
These three generators satisfy the following commutation relation,
\begin{align}
\begin{split}
[M_{11}, M_{22}]=2\,M_{22},[M_{11},M_{33}]=-2\,M_{33}, [M_{22},M_{33}]= M_{11}.
\end{split}
\end{align}
From these representations (induced on the covariance matrix) we can now identify the operators as \cite{Camargo:2018eof},
\begin{align}
\begin{split}
M_{11}\rightarrow \frac{i}{2}(\tilde x_k\,\tilde p_{k}+\tilde p_{k}\,\tilde x_k),M_{22}\rightarrow \frac{i}{2}\tilde x_k^2, M_{33}\rightarrow \frac{i}{2}\tilde p_{k}^2.
\end{split}
\end{align}
Finally, in terms of these operators we get,
\begin{align}
\begin{split}
\tilde U^{k}(\tau)=\exp\Big[\frac{i\,\rho_{k}(\tau=1)\,\tau }{2}\Big\{\sin \theta_k(\tau=0) (\tilde x_k \, \tilde p_k+\tilde p_k\, \tilde x_k)+\cos \theta_k(\tau=0) (\tilde x_k^2+\tilde p_k^2)\Big\}\Big].
\end{split}
\end{align}
Note that even though we start  with the full $GL(2, R)$ generators, the optimal circuit is composed only of the  generators belonging to the $SL(2, R)$ sub-group. 

\subsubsection*{A Brief Comparison:}

At this point, we pause and make a brief comparison between the three methods and comment on the structure of the optimal circuit. 
In all three methods, for each value of $k$ we have restricted ourselves to the space of GL(2, R) unitaries. Then we have performed an optimization to find the best possible unitary which leads to the minimal depth. Both the Fubini-Study (Section (3.1)) and covariance matrix methods (Section (3.2)) give a set of operators which satisfy SU(1,1) and SL(2, R) algebras, respectively; the optimal circuit for both these two cases are made up of scaling operators $(i\, (\tilde x_k\tilde p_k+\tilde p_k\tilde x_k))$ and $(i\,\tilde x_k^2, i\, \tilde p_k^2)$ operators. These operators are local operators in normal mode basis. On the other hand, the geodesic analysis done in the context of circuit complexity (Section (3.2)) forced us to a different set of operators ($\hat O_{k} $) as shown in (\ref{difop}), except the scaling operators ($i\, (\tilde x_k\tilde p_k+\tilde p_k\tilde x_k)$). The $\hat O_k$ operators (as mentioned in (\ref{difop})) are slightly more non-local compared to the $i\,\tilde x_k^2$ and $i\, p_k^2$ operators in the normal mode basis. We should note that when the wave function is a real  Gaussian, then we do not need any of these extra operators. Only the scaling operator $O_k$ (mentioned in (\ref{difop})) is sufficient. The expressions for complexity coming from the covariance matrix method (\ref{genCov}) and circuit complexity method (\ref{genCirc}) are basically identical, given the same reference state.  However, when the target wave function is a complex Gaussian, they are different as it is evident from (\ref{genCirc})  and (\ref{genCov}). It seems, the advantage of using Fubini-Study and covariance matrix methods is that we get the optimal circuit made from local operators, whereas the circuit complexity method tends to prefer slightly more non-local operators. However, in the next section, we will establish that the circuit complexity (from wave function) has an advantage over the other two methods as it can capture the evolution of states. 

To end this section, we stress that complexity depends on both the choice of reference state and gates (and also on the measure used). For a fixed reference state and fixed measure, the value of the complexity will depend on the underlying unitary gates. In the next section we compare the complexity obtained from the different methods---this will not be a comparison between their magnitudes, but rather a comparison of their sensitivity to a particular test that we propose.



\section{Loschmidt Echo, Fidelity, and Complexity}

In this section, we propose a diagnostic to test the advantages or disadvantages of these three different methods.  As discussed in the Introduction, we consider an interesting  
information-theoretic measure, namely the Loschmidt echo (\ref{LS}); here we discuss it in detail.

The Loschmidt echo (LE) is considered as a measure of the sensitivity of quantum mechanics to perturbations in the evolution Hamiltonian. As mentioned earlier (Eq. (\ref{LS})), the LE is defined as \cite{Echo, Echo1}\footnote{For a more comprehensive review of the application of the Loschmidt echo, interested readers are referred to \cite{Echothesis}.} 
\begin{equation} \label{LE2}
\mathcal{F}_{\text{LE}}(t)=|\langle \psi_0| \exp(i H'_1 t) \exp(-i H_1 t) |\psi_0 \rangle |.
\end{equation}
Since one is performing a
forward evolution followed by a backward evolution with slightly different Hamiltonians, the Loschmidt echo also quantifies the irreversibility in a quantum system.

\begin{figure}[ht] 
\centering
\scalebox{0.45}{\includegraphics{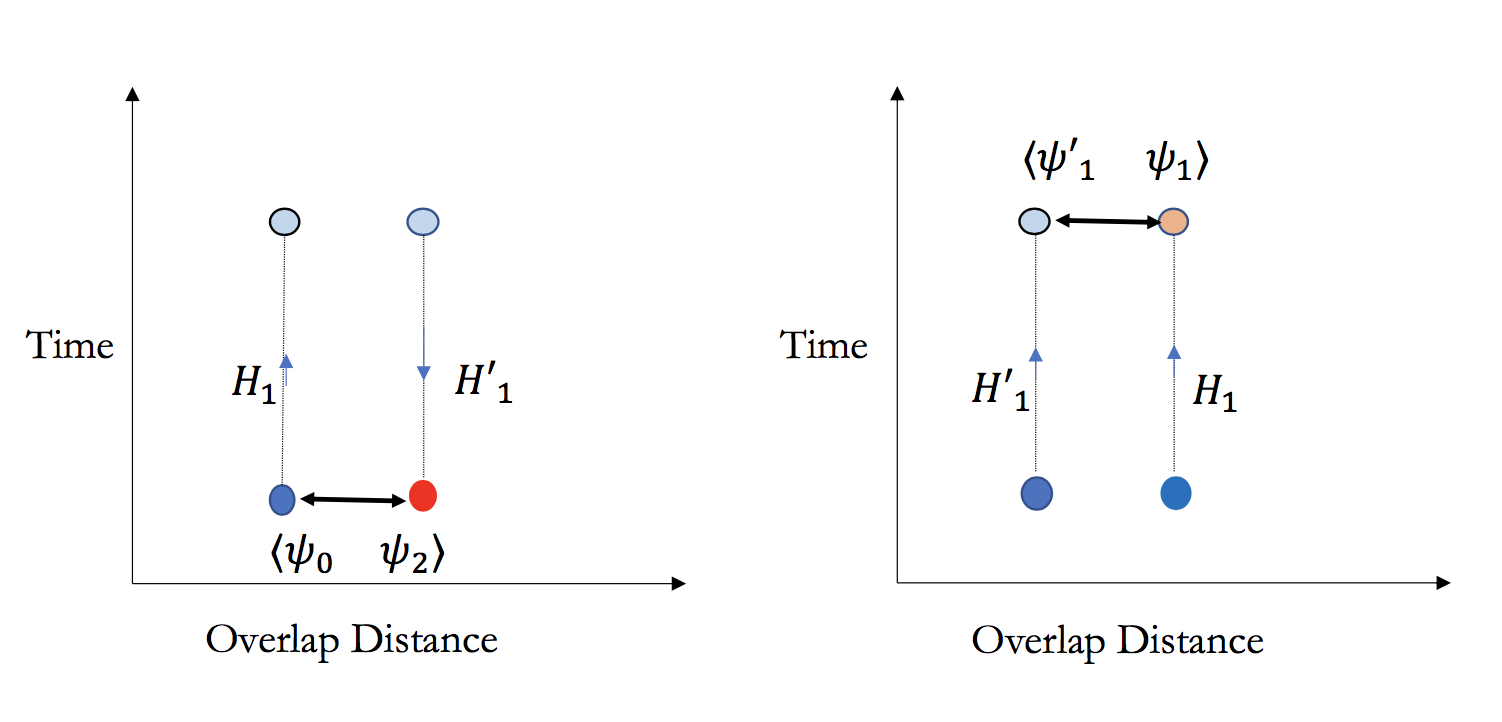} }
\caption{An illustration for Loschmidt echo and Fidelity }
\label{fig:2}
\end{figure}

For our case we take $|\psi_0\rangle$ as the ground state of the Hamiltonian at $t=0$ (Eq. (\ref{H})), which is defined in (\ref{ground}). The Hamiltonian $H_1$ (defined in Eq. (\ref{H1})) is a function of $(q_1, \hat q,).$ Also, $H'_1$ is of the form as Eq. (\ref{Hdiagonal}), however, it is a function of $(q_2,\hat q_2).$ These parameters are slightly different from both $(q, \hat q)$ and $(q_1,\hat q_1)$. We define
\be \label{wave1}
|\psi_2\rangle =\exp(i H'_1 t) \exp(-i H_1 t) |\psi_0 \rangle.
\ee
Then rewrite (\ref{LE2}) as,
\begin{equation} 
\mathcal{F}_{\text{LE}}(t)=|\langle \psi_0 |\psi_2 \rangle |.
\end{equation}
We can view (\ref{LE2}) is a different way. We define the following,
\begin{align}
\begin{split} \label{wave2}
|\psi_1\rangle= \exp(-i H_1 t) |\psi_0 \rangle,|\tilde \psi_1\rangle=\exp(-i H'_1 t)  |\psi_0 \rangle
\end{split}
\end{align}
In terms of these we can rewrite (\ref{LE2}) in the following way,
\be \label{LE3} 
 \mathcal{\tilde F}(t)=|\langle \tilde \psi_1|\psi_1\rangle |.
\ee
We termed this as \textit{Fidelity}. An illustration of Loschmidt echo and Fidelity is shown in Fig.~(\ref{fig:2}). Basically here we have defined the overlap of two wave functions evolved from the same initial state but with slightly different Hamiltonian.
Quantum mechanically (\ref{LE2}) and (\ref{LE3}) are equivalent. 
Using these overlaps, we will propose a diagnostic to distinguish between different methods of measuring complexities. We will call this {\bf `LE vs F Test'} of complexity. 
We will use the `bra' of the overlap as the reference state and `ket' as the target state. Explicitly, for Loschmidt echo we will compute the complexity of $\psi_2$ with respect to $\psi_0$ and for fidelity we will compute the complexity of $\psi_1$ with respect to $\tilde \psi_1$. Although the overlap of states-(\ref{LE2}) and (\ref{LE3}) are the same, we find that the circuit complexity (using the wave function, as done in Section~(3.2)) differs. On the other hand, complexities for Loschmidt echo and fidelity, coming from either Fubini-Study method (as in Section~(3.1)) and the circuit complexity from covariance matrix method (as in Section~(3.3)) are the same. 


\subsection{Fidelity and Complexity:} 

For Fidelity the states in the overlap are both evolved states, therefore, for this case we will compute the complexity of one evolved state with respect to the other and call this the complexity of fidelity $\mathcal{C}_{F}(\tilde U)$. Explicitly we find the complexity of evolved state $\psi_1$ by $H_1$ from $\psi_0$ with respect to $\tilde \psi_1$ by $H_1'$ from $\psi_0$. 
\begin{itemize}
\item To compute the complexity for this case using the Fubini-Study method (Section~(3.1)), we have to use the general formula mentioned in (\ref{genFS}). Now unlike (\ref{circfs}), $\theta_{1,k}$ will be non zero. In fact we will have
\begin{align}
\begin{split}
&\theta_{1,k}=2\arctanh |\gamma_{1,k}|, \theta_{2,k}=2\arctanh |\gamma_{2,k}|,\\& 
\cos(\phi_{1,k}-\phi_{2,k})=\Re\Big(\frac{\gamma_{1,k}}{\gamma_{2,k}}\frac{|\gamma_{2,k}|}{|\gamma_{1,k}|}\Big),
\end{split}
\end{align}
where $\gamma_{1,k}$ is defined in (\ref{def1}) and (\ref{def}). $\gamma_{2,k}$  is associated  with the Hamiltonian evolution of $|\psi_0\rangle$ by $H_1'$ and has the same form as $\gamma_{1,k}$ but is a function of $(q_2, \hat q_2)$ instead. These parameters are slightly different from $(q_1,\hat q_1).$
\item To compute the circuit complexity as sketched in Section (3.1) we need to use (\ref{genCirc}). Also following (\ref{tarw}) we get
\begin{align}
\begin{split} 
\tilde \psi_1(\tilde x_k,t) =\mathcal{N}(t)  \, \exp \left [- \frac{\sum_{k=0}^{N-1} \Omega_{1,k} \, \tilde x_{k}^2}{2} \right], 
\end{split}
\end{align}
where $\mathcal{N}(t)$ is the normalization and 
\be \label{omega1k}
\Omega_{1,k}=\omega_{2,k} \Big [\frac{\omega_{2,k}-i\,\omega_k \cot (\omega_{2,k}\, t)}{\omega_k-i\,\omega_{2,k}\cot (\omega_{2,k}\, t)}\Big].
\ee
Here $\omega_{2,k}$ is associated with $H_1'$ and $\omega_{2,k}^2=q_2^2+\hat q_2 \cos(\frac{2\pi\,k}{N}).$
Now in (\ref{tarw}), we need to replace $\omega_r$ by $\Omega_{1,k}.$

\item  For the covariance matrix method (Section~(3.3)) we have to use the general formula for the complexity given in (\ref{genCov}) and again replace $\omega_r=\Omega_{1,k}.$
\end{itemize}


\subsection{Loschmidt Echo and Complexity}  
 
The overlap in Loschmidt echo contains one forward and then backward evolved state and a ground state. Therefore, for this we will compute the complexity ($\mathcal{C}_{LE}(\tilde U)$) of $|\psi_2(\tilde x_k,t)\rangle$  as defined in (\ref{wave1}) w.r.t  the ground state $|\psi_0(\tilde x_k, t)\rangle$ of $H(q, \hat q)$ at $t=0.$ Now we have,
\be 
|\psi_2(\tilde x_k, t) \rangle = e^{iH'_1\, t} |\psi_1(\tilde x_k, t) \rangle,\ee
$|\psi_1(\tilde x_k,t)\rangle$ is defined in (\ref{wave1}). e have already computed it in Section~(3). We have to now do one more time evolution (backward) on it to get the $|\psi_2\rangle.$
\begin{itemize}
\item For the Fubini-Study approach, we start with the state  defined in (\ref{evolv1}).  Then we act it by $\exp(i\, H'_1\, t).$ Then again we  can decompose  $\exp(i\, H'_1\, t)$  like what we have done in (\ref{BCH}) using BCH formula with  the definitions given in (\ref{def1}). To be more explicit,
\begin{align}
\begin{split}  \label{defa}
  &\exp(i\,H'_1\, t)= \exp(\gamma_{2,k}^{+}\tau_k^{+})\exp ((\ln \gamma_{2,k}^0)\tau_k^{z})\exp(\gamma_{2,k}^{-}\tau_k^{-}),\\& \gamma_{2,k}^0=\Big(\cosh(\mu_{2,k})-\frac{\beta_{2,k}}{2\mu_{2,k}}\sinh(\mu_{2,k})\Big)^{-2},\gamma_{2, k}^{\pm}= \Big(\frac{\alpha_{2,k}^{\pm}}{\mu_{2,k}}\Big) \Big ( \frac{\sinh(\mu_{2,k})}{\cosh(\mu_{2,k})-\frac{\beta_{2,k}}{2\,\mu_{2,k}}\sinh(\mu_{2,k})} \Big ),\\&\mu_{2,k}^2=\frac{\beta_{2,k}^2}{4}-\alpha_{2,k}^{+}\alpha_{2,k}^{-}, \beta_{2,k}=i\, t\, \omega_{2,k}\, (\mathcal{\tilde U}_k^2+\mathcal{\tilde V}_k^2), \alpha_{2,k}^{+}=\alpha_{2,k}^{-}=i\,t\,\omega_{2,k} \, \mathcal{\tilde U}_k\mathcal{\tilde V}_k,\\&\mathcal{\tilde U}_{k}=\frac{\omega_{2,k}+\omega_k}{2\sqrt{\omega_{2,k}\omega_{k}}},\mathcal{\tilde V}_{k}=\frac{\omega_{2,k}-\omega_k}{2\sqrt{\omega_{2,k}\omega_{k}}}.
  \end{split}
\end{align}
Given this we need to evaluate the following,
\be
|\psi_2\rangle=\prod_{k=0}\,\mathcal{N}_{k}(t)\,\exp(\gamma_{2,k}^{+}\tau_k^{+})\exp ((\ln \gamma_{2,k}^0)\tau_k^{z})\exp(\gamma_{2,k}^{-}\tau_k^{-})\exp(\gamma_{1, k}^{+}(t)\, a^{\dagger}_{k}a^{\dagger}_{-k})|k,-k\rangle,
\ee
We know that $\tau^{-}_k$ annihilates $|k,-k\rangle$. Then we  successively use BCH formula and the decomposition mentioned in (\ref{BCH}). Finally, after absorbing overall phase factors inside the normalization we get,
\be \label{finalpsi}
|\psi_2\rangle=\prod_{k=0}\mathcal{\tilde N}_{k}(t)\exp(\hat \gamma_{k}\, \tau_{k}^{+})|k,-k\rangle
\ee
where
\be \label{finalfactor}
\hat \gamma_k=\gamma_{2,k}^{+}+\frac{\gamma_{1,k}^{+}\gamma_{2,k}^{0}}{1-\gamma_{1,k}^{+}\gamma_{2,k}^{-}}.
\ee
 Then we compute the complexity for this case. We can simply use the formula (\ref{circfs}) and replace $\gamma_{1,k}$ by $\hat \gamma_k$ as mentioned in (\ref{finalfactor}).

\item  For computing complexity by the methods outlined in Section~(3.2) and Section~(3.3), we need the following evolved state $\psi_2(x,t),$
\bea
\psi_2(\tilde x_k,t) &= &\prod_{k=0}^{N-1}\mathcal{\hat N}_{k}(t) \exp \left[ -\frac{1}{2}\hat \Omega_{k} \tilde x_k^2   \right] ,
\eea
where \be\label{omedef} \hat \Omega_{k}=  i \ \omega_{2,k} \cot \omega_{2,k} t  + \frac{\omega_{2,k}^2}{  \sin^2 \omega_{2,k} t (\Omega_{k} + i \omega_{2,k} \cot \omega_{2,k} t)} \ee and $\mathcal{\hat N}_{k}(t)$ is the normalization factor so that the inner-product of the wave function with itself remains one.  $\omega_{2,k}$ is defined below equation (4.8). To compute the complexity using either of these two methods we can simply use either (\ref{circ}) or (\ref{comCov}) and replace $\Omega_{k}$ by $\hat \Omega_{k}$ as defined in (\ref{omedef}).
\end{itemize}


\subsection{LE vs F Test for Different Methods of Complexities}

Now we explicitly evaluate $\mathcal{C}_{\text{F}}(\tilde U)$ and $\mathcal{C}_{\text{LE}}(\tilde U)$ coming from three different methods. This will show the difference between these three method, namely  we will get, $\mathcal{C}_{\text{F}}(\tilde U)=\mathcal{C}_{\text{LE}}(\tilde U)$ for Fubini-Study and covariance matrix method but $\mathcal{C}_{\text{F}}\neq \mathcal{C}_{\text{LE}}(\tilde U)$ from circuit complexity method as described in Section~(3.2). We evaluate all the expressions and present two representative plots  by choosing the following values for the parameters,
\be \label{paravalue}
 \{q^2=5,q_1^2=20,q_2^2=29, \hat q=4,\hat q_1=16,\hat q_2=-20\}.
  \ee
  with two choices for for $N=500$ and $N=1000.$

\subsubsection*{Complexity from Fubini-Study:}
For this case we have the following expressions for the complexities-
\begin{align}
\begin{split} \label{diff1}
&\mathcal{C}_{\text{LE}}(\tilde U)=\sqrt{\sum_{k=0}^{N-1} (\arctanh |\hat \gamma_{k}|)^2},\\&
\mathcal{C}_{\text{F}}(\tilde U)=\frac{1}{2} \sqrt{\sum_{k=0}^{N-1}\left(\arcosh\Big[\cosh (\theta_{1,k})\cosh(\theta_{2,k})-\sinh(\theta_{1,k})\sinh(\theta_{2,k})\Re\Big(\frac{\gamma_{1,k}}{\gamma_{2,k}}\frac{|\gamma_{2,k}|}{|\gamma_{1,k}|}\Big)\Big] \right)^2},
\end{split}
\end{align}
where $ \theta_{1,k}=2\arctanh |\gamma_{1,k}|, \ \theta_{2,k}=2\arctanh |\gamma_{2,k}|$ and $\gamma_{1,k}$ and $\hat \gamma_{k}$ are defined in (\ref{def}) and (\ref{finalfactor}) respectively. \\
\begin{figure}[ht]
\centering
  \scalebox{0.60}{\includegraphics{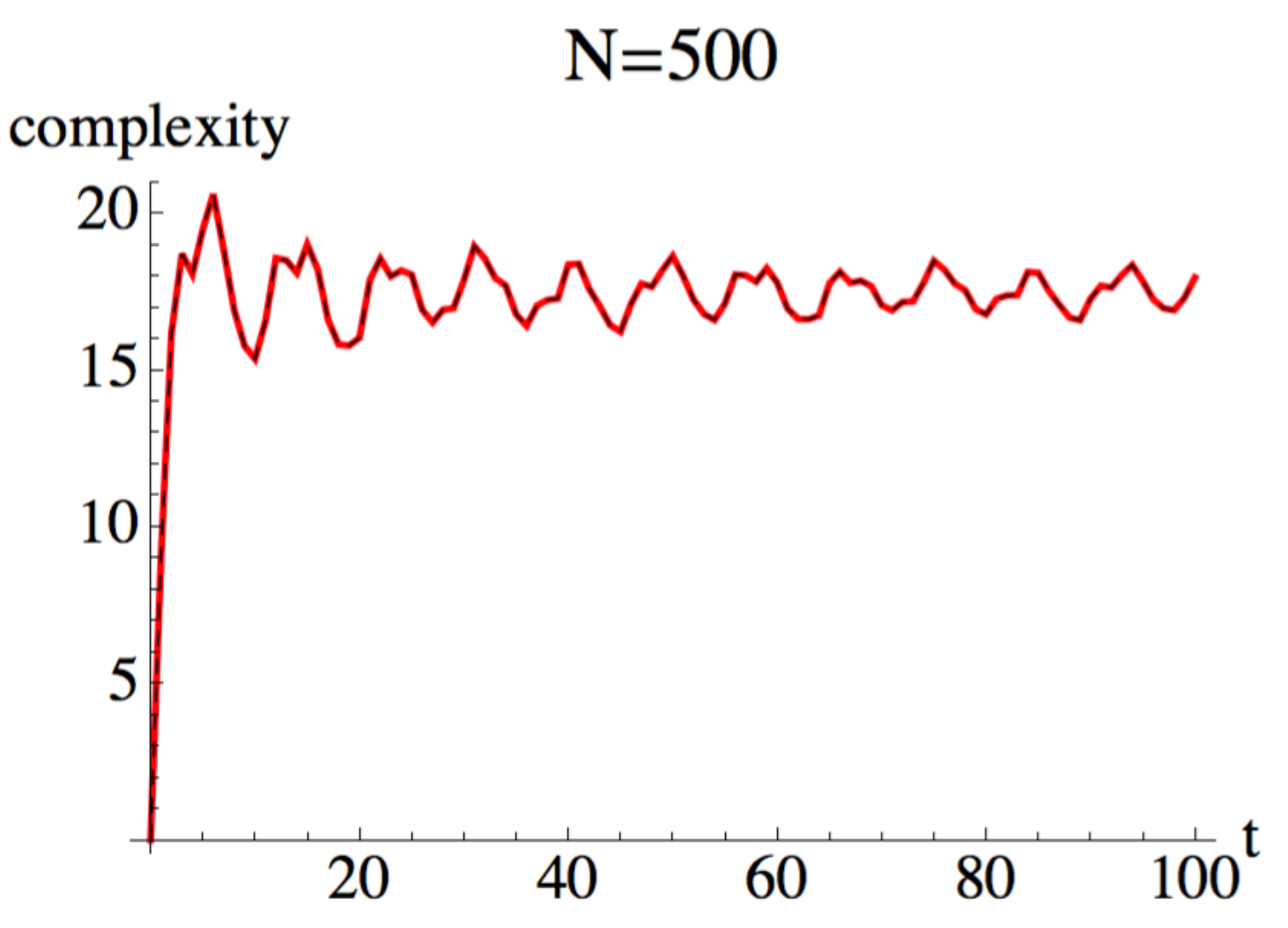}  }
  \scalebox{0.60}{\includegraphics{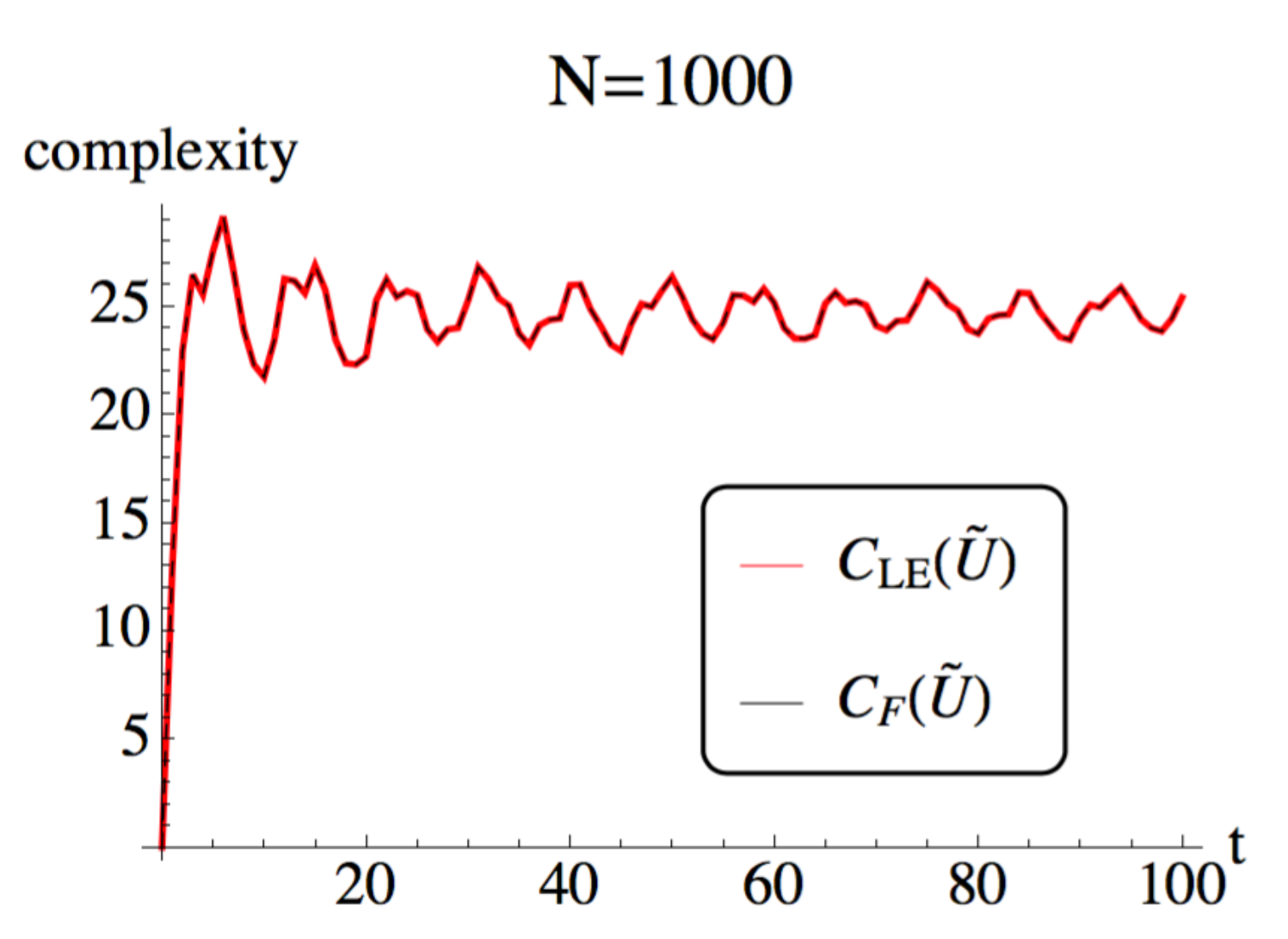}  }
\caption{LE vs F Test  for Fubini-Study}
\label{FB}
\end{figure}

\noindent
From Fig.~(\ref{FB}) one can immediately see that the Fubini-Study approach cannot distinguish between the two complexities. They overlap with each other completely. At this point we are unable to prove analytically that the two expressions $\mathcal{C}_{\text{LE}}(\tilde U)$ and $\mathcal{C}_{\text{F}}(\tilde U)$ mentioned in (\ref{diff1}) are equal to each other.  However, we can only show that they are equal, at the leading order in small $t$ expansion. We sketch the proof in the appendix~(\ref{AppC}) and leaving the complete proof for future study. 

\subsubsection*{Circuit Complexity:} 
For this case we have the following,

\bea
\mathcal{C}_{\text{LE}} (\tilde U) &= & \frac{1}{2}\sqrt{ \left[ \sum_{k=0}^{N-1}\Big(\log \frac{ |\hat \Omega_{k} | }{\omega_{k} } \Big)^2+ \Big(\arctan \frac{\text{Im}(\hat \Omega_{k}) }{\Re (\hat \Omega_{k})}\Big )^2 \right ]}, \cr
\mathcal{C}_{\text{F}} (\tilde U) &=& \frac{1}{2}\, \sqrt{
 \sum_{k=0}^{N-1} \left[\Big(\log\frac{|\Omega_k|}{|\Omega_{1,k}|} \Big)^2+ \Big(\arctan\frac{\Re(\Omega_{1,k})\text{Im}(\Omega_k)-\Re(\Omega_k)\text{Im} (\Omega_{1,k})}{\Re(\Omega_{1,k})\Re(\Omega_k)+\text{Im}(\Omega_k)\text{Im}(\Omega_{1,k})}\Big)^2 \right ]},
\eea
where, $\Omega_k$ and $\Omega_{1,k}$ are defined in (\ref{omegak}) and (\ref{omega1k}) respectively.\\

\noindent

\begin{figure}[ht] 
\centering
 \scalebox{0.60}{\includegraphics{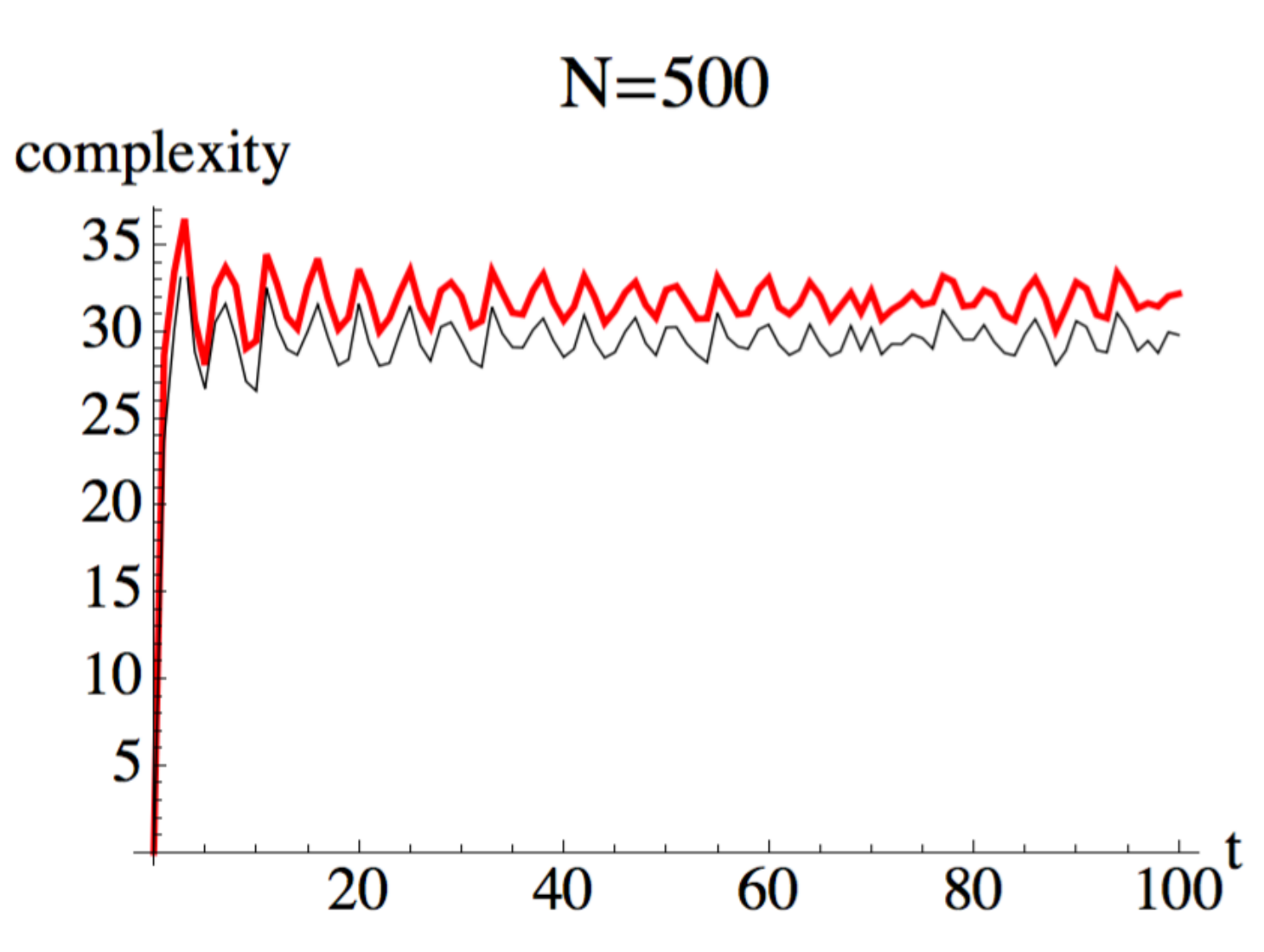} }
 \scalebox{0.60}{\includegraphics{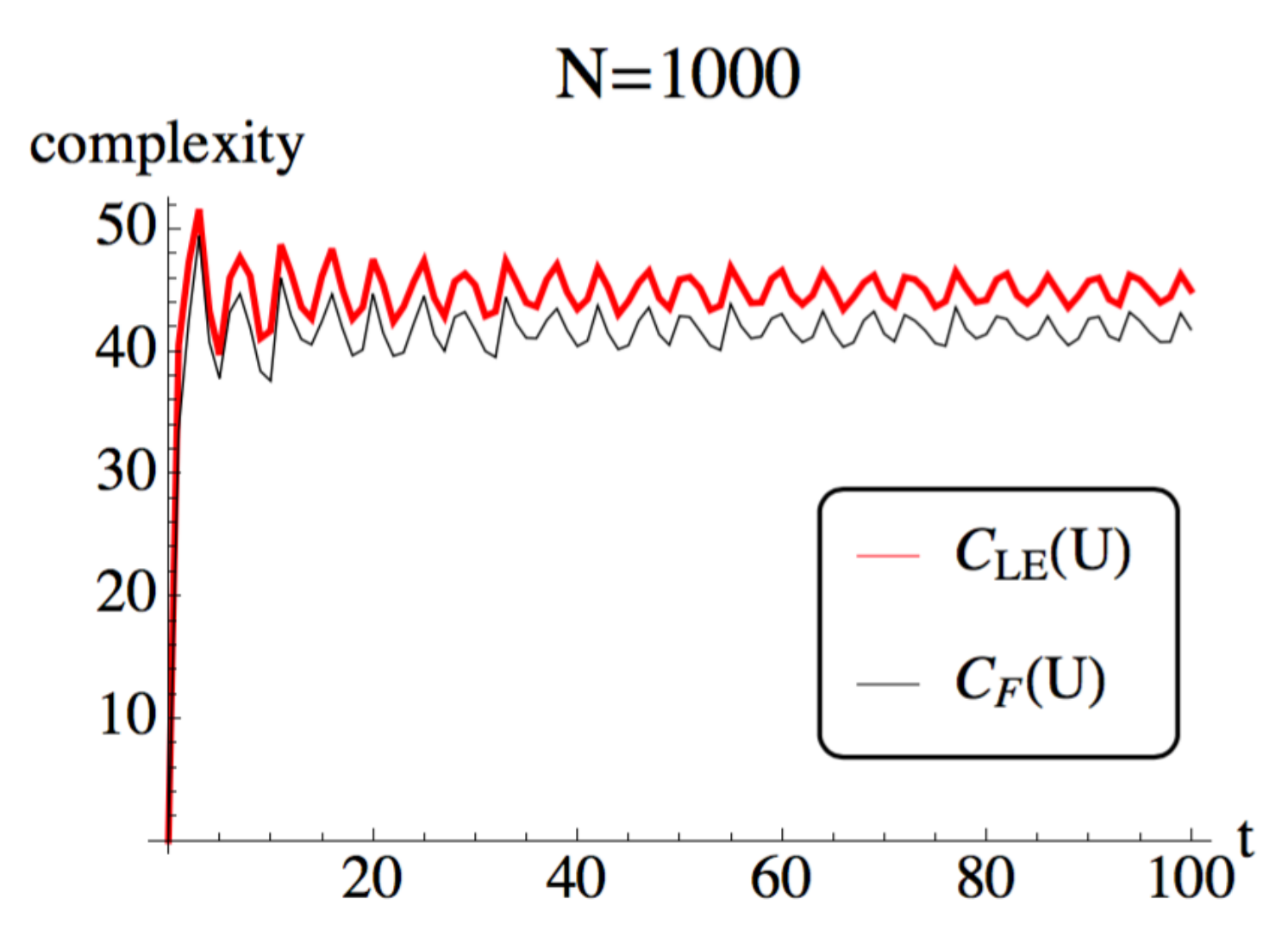} }
\caption{LE vs F Test for Circuit Complexity}
\label{CC}
\end{figure}
\begin{figure}[ht]
\centering
  \scalebox{0.60}{\includegraphics{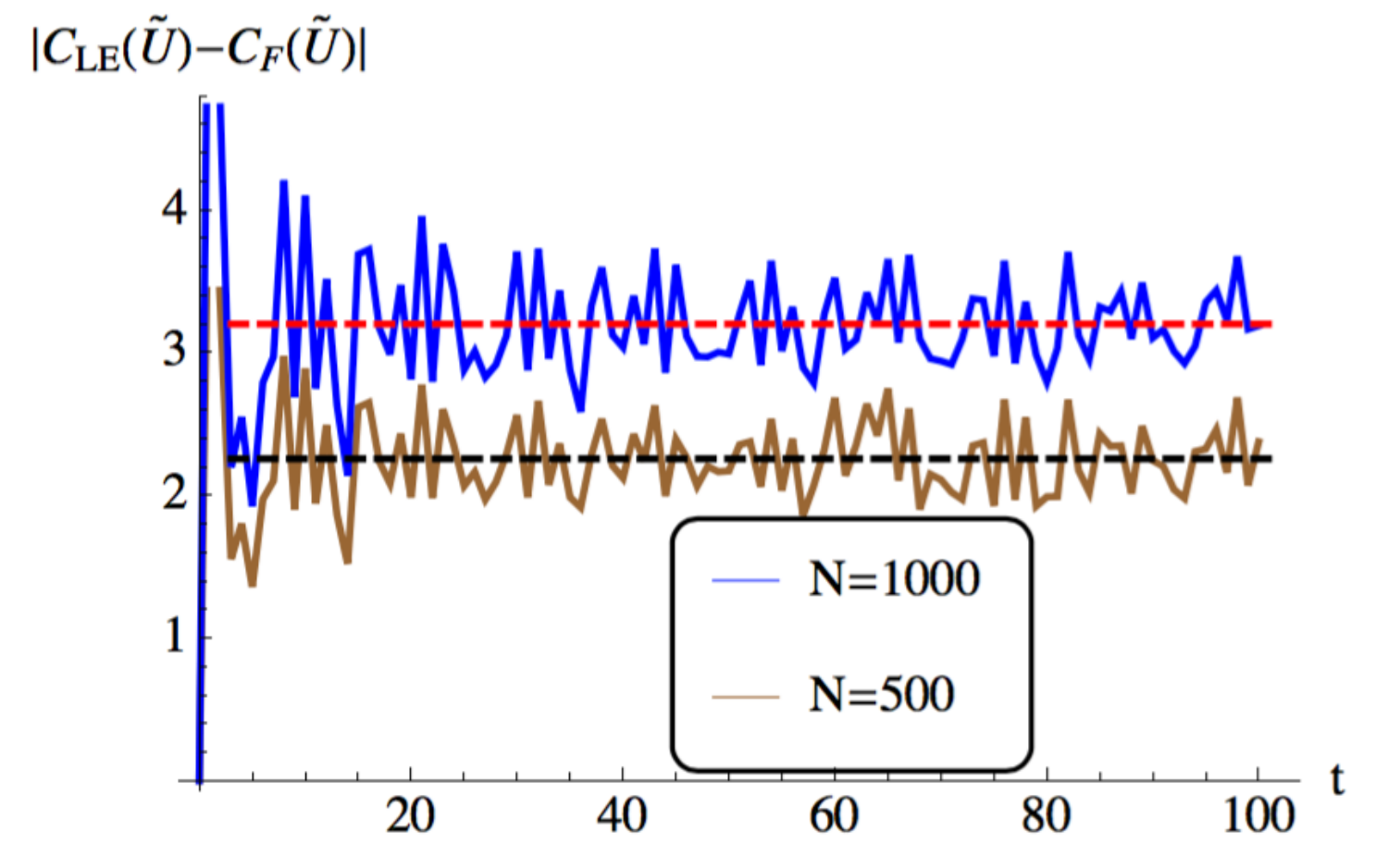} }
\caption{Difference plot for circuit complexity  method. Dashed lines corresponds to the value around which $|\mathcal{C}_{\text{LE}}(\tilde U)-\mathcal{C}_{\text{F}}(\tilde U)|$ fluctuates at late times which are  approximately, 3.2 ( for N=1000, corresponding to the  red dashed line) and 2.2 (for N=500, corresponding to the black dashed line) respectively. }
\label{Diff2}
\end{figure}

From Fig.~(\ref{CC}) it is evident that the two complexities are quite different in circuit complexity method. Moreover, the complexity related to Loschmidt Echo (complexity for the state, we obtained by a forward followed by a backward evolution) is larger than the complexity that we get for the Fidelity, namely the complexity between two forward evolved state for most part of the evolution.
\bea \label{conclude}
|\langle \psi_0 |\psi_2 \rangle| &=& |\langle \tilde \psi_1 |\psi_1 \rangle| \cr
\mathcal{C} (\psi_2 , \psi_0) &>& \mathcal{C} (\psi_1, \tilde \psi_1) 
\eea
Therefore, although the closeness of states between ($\psi_0$ and $\psi_2$) is same as the closeness between ($\tilde \psi_1$ and $\psi_1$), the complexity of $\psi_2$ with respect to $\psi_0$ is larger than the complexity of  $\psi_1$ with respect to $\tilde \psi_1$. We further plot $|\mathcal{C}_{\text{LE}}(\tilde U)-\mathcal{C}_{\text{F}}(\tilde U)|$ with respect to time. From Fig.~(\ref{Diff2}) we observe that this difference becomes constant quite fast and just fluctuates around this constant value even at large time. It will be interesting to do further numerical analysis to explore the late time behaviour and investigate their physical implications in a future work.

So far we have only used, $\mathcal{F}_2$ as the measure for the complexity. We have also considered another measure of complexity, namely $\mathcal{F}_{\kappa=1},$ as defined in (\ref{meas}) with $p_I=1$ for all $I$. Using the arguments of \cite{recentmyers} one can easily show that, like $\mathcal{F}_{2}$, $\mathcal{F}_{\kappa=1}$ also get minimized when evaluated on the same geodesic solution (\ref{Newsol1}). Then we can numerically show that the complexities associated with Loschmidt Echo and Fidelity are different with respect to this $\mathcal{F}_{\kappa=1}$ measure.

\subsubsection* {Complexity from the covariance matrix:} For this case we have the following,
\begin{align}
\begin{split} \label{diff2}
&\mathcal{C}_{\text{LE}}(\tilde U)= \frac{1}{2} \sqrt{ \sum_{k=0}^{N-1} \left( \arcosh \left( \frac{\omega_k^2+ |\hat \Omega_k |^2 }{2 \  \omega_k \Re (\hat \Omega_k)}\right) \right)^2 },\\&
\mathcal{C}_{\text{F}}(\tilde U)=\frac{1}{2}\sqrt{\sum_{k=0}^{N-1}\left(\arcosh \left(\frac{\Re(\Omega_{1,k})^2+\Re(\Omega_k)^2+(\text{Im}(\Omega_{1,k})-\text{Im}(\Omega_k))^2}{2\,\Re(\Omega_{1,k}) ,\Re(\Omega_k)}\right)\right)^2},
\end{split}
\end{align}
where, $\Omega_k$ and $\Omega_{1,k}$ are defined in (\ref{omegak}) and (\ref{omega1k}) respectively.\\
\begin{figure}[ht]
\centering
 \scalebox{0.60}{\includegraphics{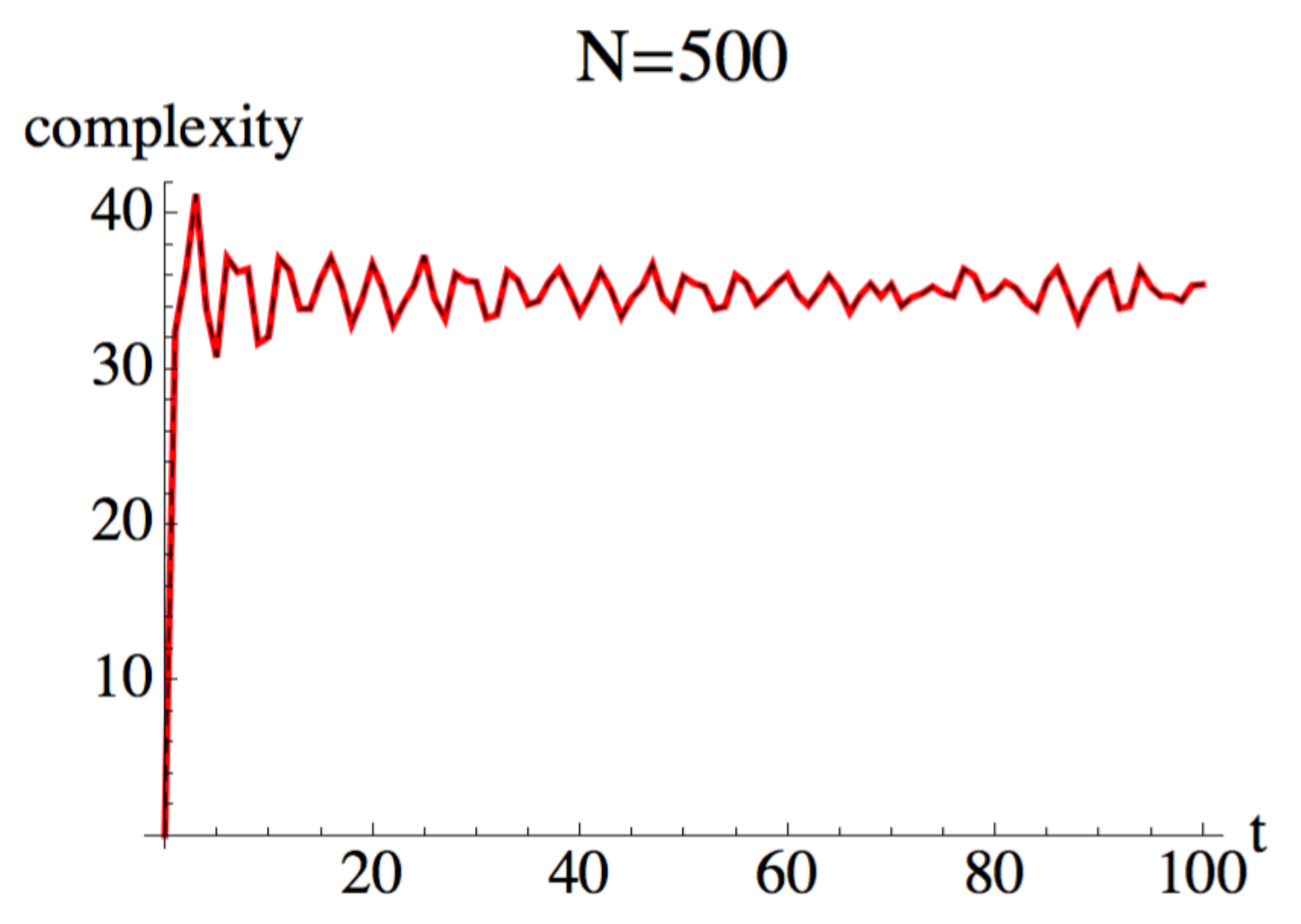}  }
 \scalebox{0.60}{\includegraphics{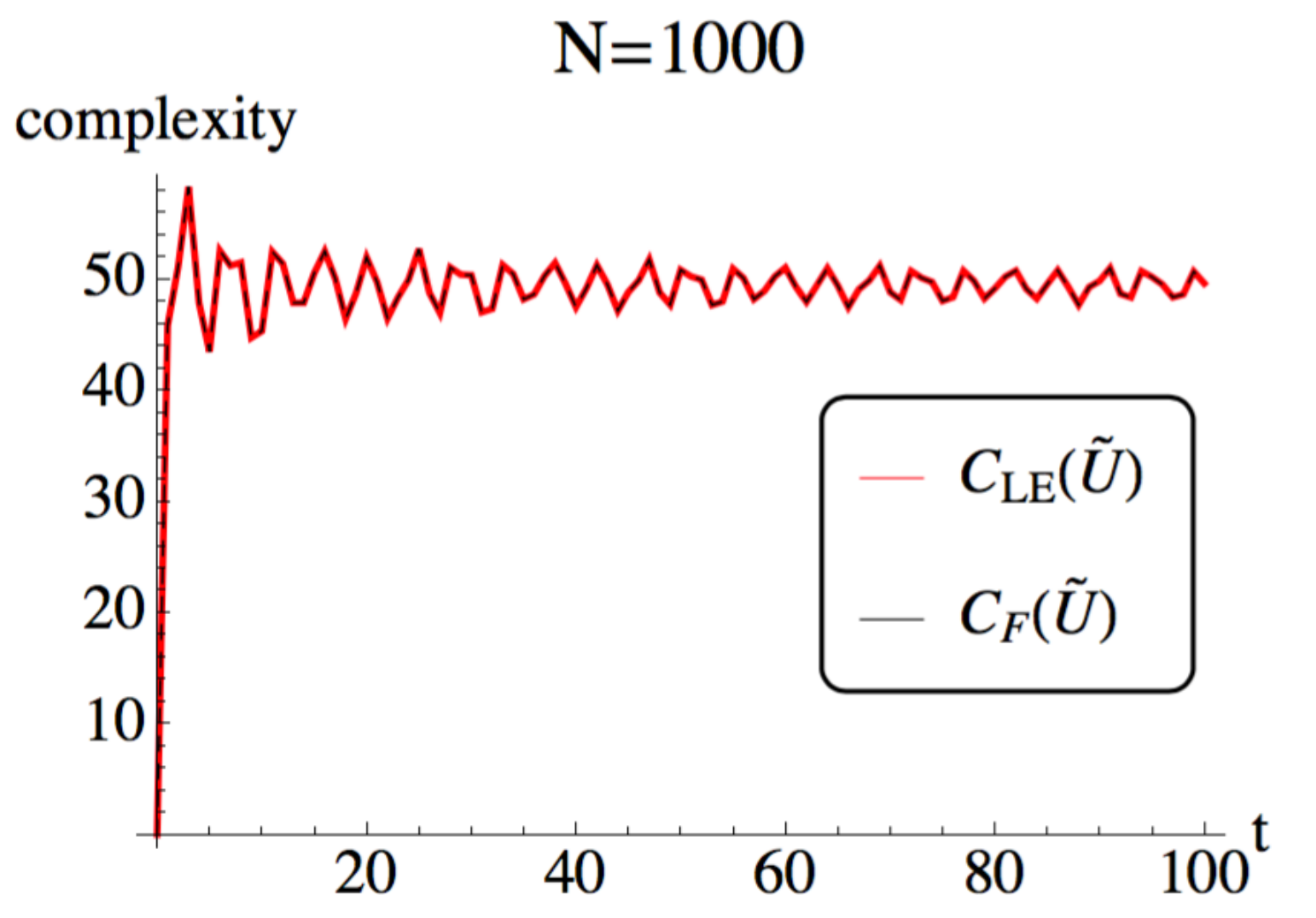}  }
\caption{LE vs F Test for covariance matrix Method (Section~(3.3))}
\label{CCC}
\end{figure}

\noindent
Just like the Fubini-Study approach, from Fig.~(\ref{CCC}) one can immediately see that the covariance matrix methods cannot distinguish between the two complexities. They overlap with each other completely and again this behaviour is independent  of the values of the parameters and $N.$ Again, like the Fubini-Study  case, we are unable to prove analytically that these two expressions $\mathcal{C}_{\text{LE}}(\tilde U)$ and $\mathcal{C}_{\text{F}}(\tilde U)$ mentioned in (\ref{diff2}) are equal to each other. However, we can only show that they are equal at the leading order in small $t$ expansion. We sketch the proof in the appendix~(\ref{AppC}) and leaving the complete proof for future study. 
\par

Now we would like to stress the following point. In the  literature, for example, \cite{Echo,Echo1, Echothesis}, the Loschmidt echo is considered as a diagnostic for quantum chaos and for that it is imperative to consider the two Hamiltonians $H_1(q_1,\hat q_2)$ and $H_1'(q_2,\hat q_2)$ are only slightly different. In our notation this corresponds to the  fact that the values of the parameter sets $\{q_1,\hat q_1\}$ and $\{q_2,\hat q_2\}$ are slightly different from each other.  However the results presented in this section and the next, only require that the values of these two sets of parameters have to be different from each other, but  they do not depend on the magnitude of this difference. In fact  difference between $\{q_1,\hat q_1\}$ and $\{q_2,\hat q_2\}$ can be large and the choice made in (\ref{paravalue}) corroborate this statement.
\subsubsection*{A Consistency Check}
\par
\begin{figure}[ht] 
\centering
\scalebox{0.65}{\includegraphics{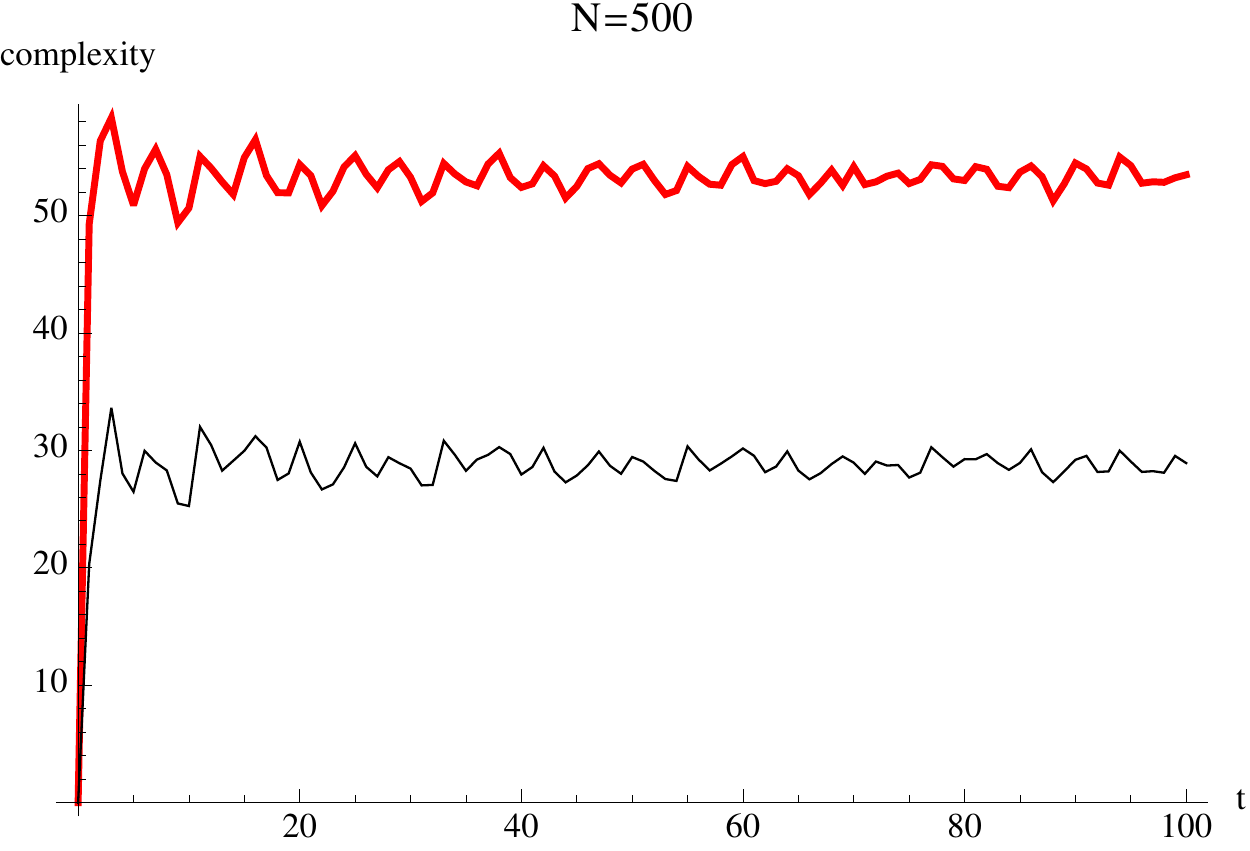}} 
 \scalebox{0.65}{\includegraphics{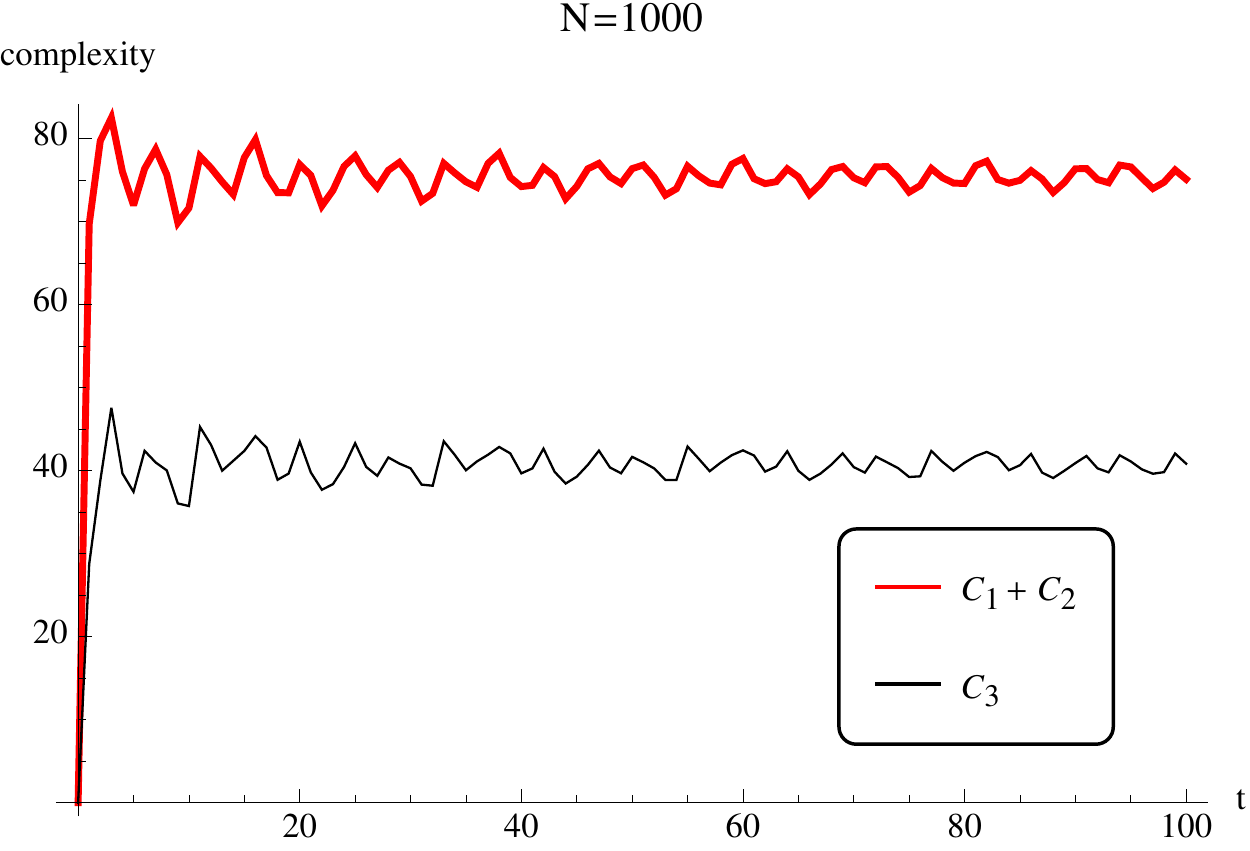}} 
\caption{Triangle Inequality for Circuit Complexity}
\label{triangle}
\end{figure}
We will conclude this section with a quick check of the triangle
inequality. This also serves as a consistency check for our computation. We considered states-$\{\psi_0, \psi_1,\psi_2\}$ and computed circuit complexities (Section~(3.1)) for the state $\psi_1$ (forward evolved by $H_1$) with respect to $\psi_0$ ($\mathcal{C}_1$), the state $\psi_2$ (forward evolution by $H_1$ followed by a backward evolution by $H'_1$) with respect to $\psi_0$ ($\mathcal{C}_2$) and finally for the state $\psi_2$ with respect to $\psi_1$ ($\mathcal{C}_3$). $\psi_0$ is the ground state of the Hamiltonian (\ref{H}). Then the Fig.~(\ref{triangle}) clearly shows that the triangle inequality ($\mathcal{C}_1+\mathcal{C}_2\,\geq\, \mathcal{C}_3$) is satisfied. For the other two methods (in Section~(3.1) and Section~(3.2)) we can check that triangle equality is trivially satisfied. This is a consistency check for our numerical computations. 


\section{A Generic Feature of Circuit Complexity}


In the previous section, we established that only circuit complexity is sensitive to the evolution of states; in that sense, one can argue that it is a better measure of complexity. 
Now we will explore if this is a generic feature of the overlap, namely if we do further forward and backward evolutions by another set of Hamiltonians $H_2$ and slightly different $H'_2$. 
\begin{figure}[ht] 
\centering
 \scalebox{0.65}{\includegraphics{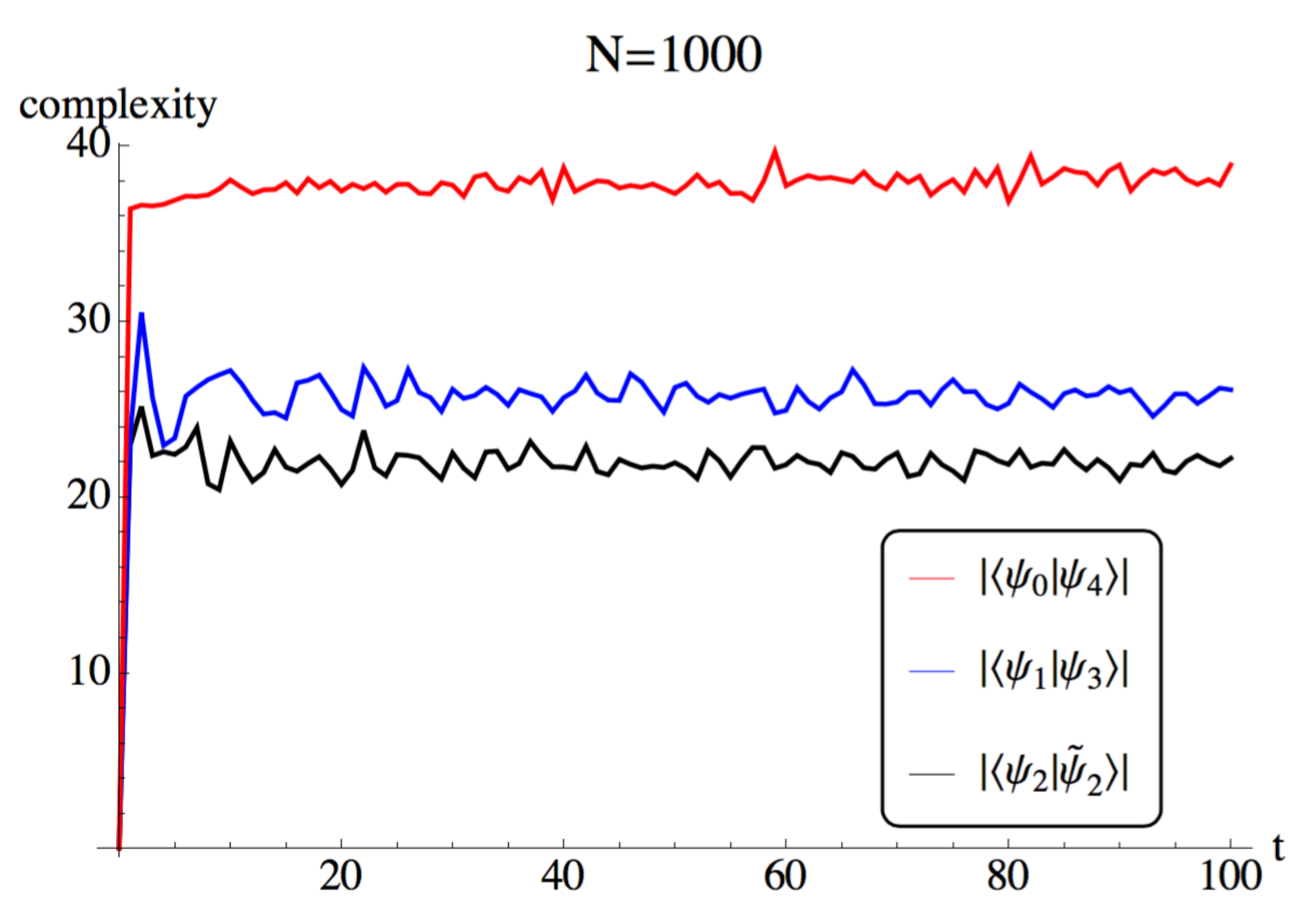}  }
\caption{Evolution with four Hamiltonians. We have used the following set of parameters :$\{q_0^2=5,\hat  q_0=4\}$ for $H_0 ( \rm{for}\,, t \leq 0)$ and $\{q_1^2=20,\hat q_1=16,q_2^2=34,\hat q_2=-30,q_3^2=117,\hat q_3=108,q_4^2=208,\hat q_4=-192\}$ for $H_1,H_1', H_2, H_2'$ respectively.}
\label{4ham}
\end{figure}
\be \label{4evolve}
|\psi_4(\tilde x_k, t) \rangle =  e^{iH'_2 t} e^{-i H_2 t}e^{iH'_1 t} e^{-i H_1 t} |\psi_0\rangle
\ee
The analogue of Loschmidt echo will be 
\be \label{overlap}
\mathcal{F}_{\text{LE}}= \langle \psi_4 | \psi_0\rangle
\ee
Note that this overlap (\ref{overlap}) can be written as follows:
\be
\langle \psi_4 | \psi_0\rangle =  \langle \psi_1 | \psi_3\rangle 
=\langle \tilde \psi_2 | \psi_2\rangle,
\ee
where 
\bea
|\psi_3(\tilde x_k, t) \rangle &=&  e^{-i H_2 t}e^{iH'_1 t} e^{-i H_1 t} |\psi_0\rangle,  |\psi_1(\tilde x_k, t) \rangle = e^{-i H'_2 t} |\psi_0\rangle\,, \cr
|\psi_2(\tilde x_k, t) \rangle &= & e^{iH'_1 t} e^{-i H_1 t} |\psi_0\rangle, |\tilde \psi_2(\tilde x_k, t) \rangle =  e^{i H_2 t} e^{-i H'_2 t} |\psi_0\rangle\,.
\eea
Therefore, we get two different overlaps with the same magnitude. Hence we get two different types of Fidelities in terms of states involved in the overlap. We will label $ \langle \psi_1 | \psi_3\rangle$  as Fidelity $\mathcal{F}_1$ and $\langle \tilde \psi_2 | \psi_2\rangle$ as Fidelity $\mathcal{F}_2$. Note that, one can also consider the reverse combination of states, when defining Fidelity, such as $\langle \psi_2 | \tilde \psi_2\rangle$ and $\langle \psi_3 | \psi_1\rangle$. These extra overlaps will not change the qualitative feature of our results, therefore, for illustration purposes we will ignore them. After performing the appropriate number of evolutions, we compute the corresponding complexities. Our computation is summarized in Fig.~(\ref{4ham}). Once again we see that the complexity for the LE is always larger than complexity computed for any combinations of intermediate states corresponding to Fidelity. This result can be written as 
\be \label{sr}
\mathcal{C}_{\text{LE}}^{(\psi_4 , \psi_0)}(\tilde U) > \mathcal{C}_{\text{F}_1}^{(\psi_3 , \psi_1)}(\tilde U) >  \mathcal{C}_{\text{F}_2}^{(\psi_2 , \tilde \psi_2)}(\tilde U) 
\ee
The superscripts denote the pair of wave functions for which we compute the complexity.\par
 Now there are several comments are in order.
\begin{itemize}
\item We made an important assumption here that, each evolution (forward and backward) are done by a different Hamiltonian. Let us try to elaborate this point. For the case of LE ($\langle \psi_0 | \psi_4\rangle $) in the above example, $|\psi_4\rangle$ is being generated from  $|\psi_0\rangle$ by 4 evolutions by 4 different Hamiltonians \footnote{We thank the referee for raising this point.}. One can obviously generalize this argument for any number of evolutions. Now once we fix the set of the Hamiltonians entering in LE, we can easily compute various types of Fidelity by distributing this same set of Hamiltonians in various different ways such that quantum mechanically all of these overlaps will be the same. For the case stated earlier, there will be two distinct types of fidelities namely,  $\langle \psi_1 | \psi_3\rangle$ and $\langle \tilde \psi_2 | \psi_2\rangle $. 
\item Another point is that, for the purpose of these computations our starting point is always the ground state of the Hamiltonian (for example $|\psi_0\rangle$ in the equation (\ref{4evolve})) which is of the form (\ref{hamilton}). All the other states are constructed by evolving this ground state.

\item{ Last but the not the least, the result in (\ref{sr}) does not depend on whether we are performing a forward or a backward evolution and also does not depend on the degree to which these Hamiltonians differ from each other.}
\end{itemize}

\begin{figure}[ht] 
\centering
 \scalebox{0.50}{\includegraphics{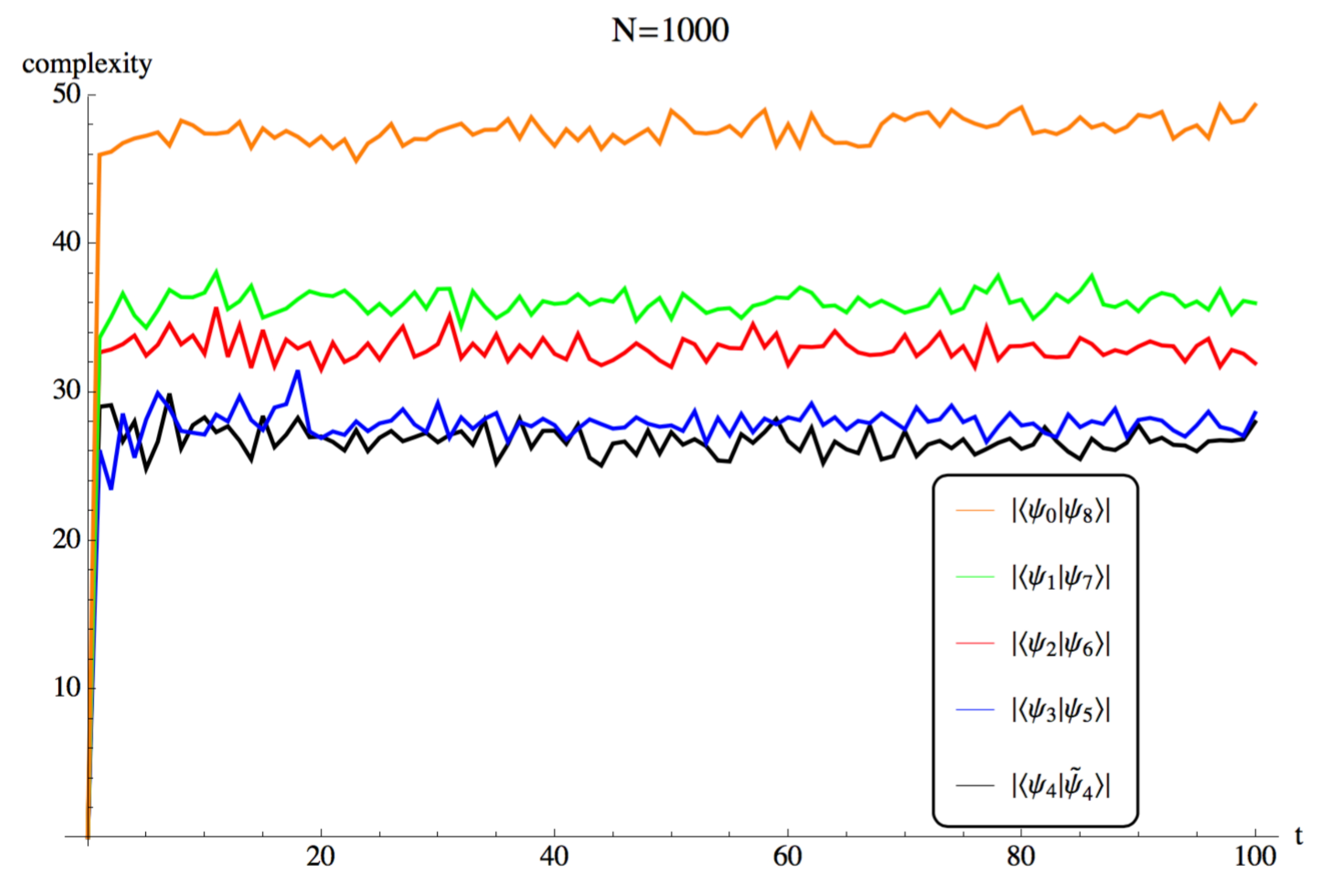}  }
\caption{Evolution with eight Hamiltonians. We have used the following set of parameters :$\{q_0^2=5,\hat  q_0=4\}$ for $H_0 ( \rm{for}\,, t \leq 0)$ and $\{q_1^2=20,\hat q_1=16,q_2^2=34,\hat q_2=-30,q_3^2=117,\hat q_3=108,q_4^2=208,\hat q_4=-192,q_5^2=325,\hat q_5=300,q_6^2=468,\hat q_6=-432, q_7^2=637,\hat q_7=588,q_8^2=832,\hat q_8=-768\}$ for $H_1,H_1', H_2, H_2',H_3, H_3', H_4, H_4'$ respectively.}
\label{8ham}
\end{figure}

Now given these facts we next try to generalize our results. We can perform the same operations for arbitrary number of different Hamiltonians, leading to arbitrary number of evolutions for the state. We have tested this for 8 different evolutions with different Hamiltonians and interestingly, we find that the complexity corresponding to Loschmidt echo is always larger than any possible fidelity. Moreover, for different fidelities, the number of evolutions performed on reference state dictates the magnitude of their complexities.  This is shown in the Fig~(\ref{8ham}). 

This result motivate us to make the following conjecture-\\ 

\noindent
{\it Overlap with the largest number of evolutions acting on the same reference state always corresponds to the highest complexity. }
\be
\mathcal{C}_{\text{LE}}^{ (\psi_n , \psi_0)}(\tilde U) > \mathcal{C}_{\text{F}_1}^{(\psi_{n-1} , \psi_1)}(\tilde U) > \mathcal{C}_{\text{F}_2}^{(\psi_{n-2} , \psi_2)}(\tilde U) > .. ... >  \mathcal{C}_{\text{F}_n}^{(\psi_{n/2} , \tilde \psi_{n/2})}(\tilde U). 
\ee

This result implies, although the closeness between two states (overlap) does not change with unitary evolutions, 
they are very different from the perspective of a quantum circuit and the difficulty (in terms of complexity) of getting an evolved target state from the other. And our analysis also gives
us a guideline about pairs for which it is easier to move between states in the sense of complexity. It tells us which pair of states will have the smallest complexity for a given set of Hamiltonians.

One interesting feature of these differences in complexity for the overlaps is that they do not die away with time. The differences are small at very early time, but it become fixed as soon as the complexities saturate.
Moreover, our analysis indicates an upper bound on the complexity for a given overlap evolved with a fixed number of Hamiltonians. Since as overlaps they are all the same, this comparison by complexity can be seen as comparing the same quantity from different ways, therefore, by abusing the language we can say that, any pair other than the pair with maximum complexity will have uncomplexity or resources \cite {Hol8}. In our language, \\

\noindent
{\it The complexity corresponding to the Loschmidt Echo will always have the highest complexity. Therefore, the complexity corresponding to the Fidelity will have resources.} \\

Given the recent experimental advances one can possibly simulate these overlaps in an experimental setting \cite{Echo, Swingleex, Swingleex1}. Our complexity analysis is providing us with the most efficient (optimal) quantum circuit needed to simulate the time evolution, 
hence providing a natural selection mechanism. Note that quantum mechanically all of these overlaps are the same. Therefore, this test will reduce the difficulties of experimental implementation in the sense that it can be obtained by constructing a quantum circuit with a minimal number of gates. \par  
Before ending this section, we want to further clarify what we meant by having \textit{resources}. Again, let's assume that we want to make overlap measurements (say, for two steps of evolutions) in the lab. We can measure either Loschmidt Echo or Fidelity since  quantum mechanically the result is the same. Now suppose we are being \textit{supplied} with either $\psi_1$ or $\tilde \psi_1,$ apart form $\psi_0$ then our result implies that it is easier to measure Fidelity as the complexity between the states entering in Fidelity is less compared to the complexity between the states entering in Loschmidt Echo. Note that, if we do not have either of $\psi_1$ or $\tilde \psi_1,$ but only $\psi_0$ then of course we will loose this advantage since then we have to also take into account the complexity of simulating $\psi_1$ or $\tilde \psi_1$ from $\psi_0.$ We will have this advantage only  when we are being supplied with either  $\tilde \psi_1$ or $\psi_1.$ This is precisely what we meant by having \textit{resources} i.e having possession of states with some extrinsic complexity (w.r.t $\psi_0$). 



\section{Evolution of Complexity: Local {\it vs} Non-Local Theory}


In this section, we explore the evolution of complexity in a different context; to highlight the point, we show results obtained from the circuit complexity method (Section (3.2)), but our discussion is applicable to the other two methods as well (discussed in Sections  (3.1) and  (3.2)).

As seen in Fig.~(\ref{CC}), $\mathcal{C}_{\text{LE}}(\tilde U)$ and $\mathcal{C}_{\text{F}} (\tilde U)$ grow almost instantaneously, and then fluctuate around a constant value.\footnote{This behavior was observed for all values of the parameters considered.} While the fluctuations are not unexpected\cite{Hol8}, the fast growth is not in conformity with some of the expectations in the existing literature \cite{Hol8}.
Furthermore, we have found the complexity attains saturation faster than the entanglement entropy \cite{ab, WP}.
Although it is a bit early to do a direct comparison with Holography, we note that this feature contradicts the holographic expectations, namely that the complexity grows more slowly than the entanglement and attains saturation much later. 
We will address this issue of the different time scales in upcoming works \cite{ab, WP}.

\begin{figure}[ht] 
\centering
 \scalebox{0.39}{\includegraphics{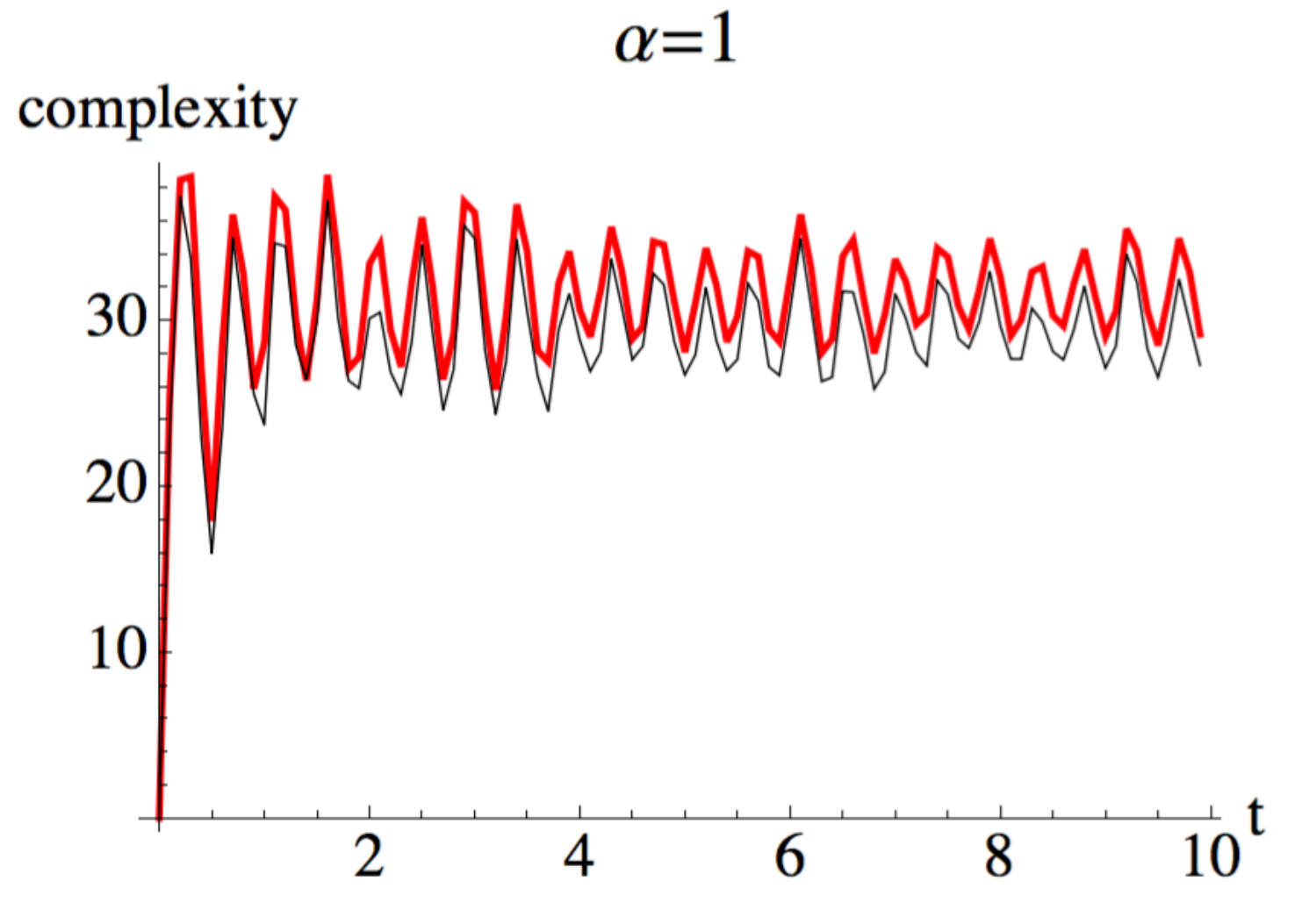}}
 \scalebox{0.39}{\includegraphics{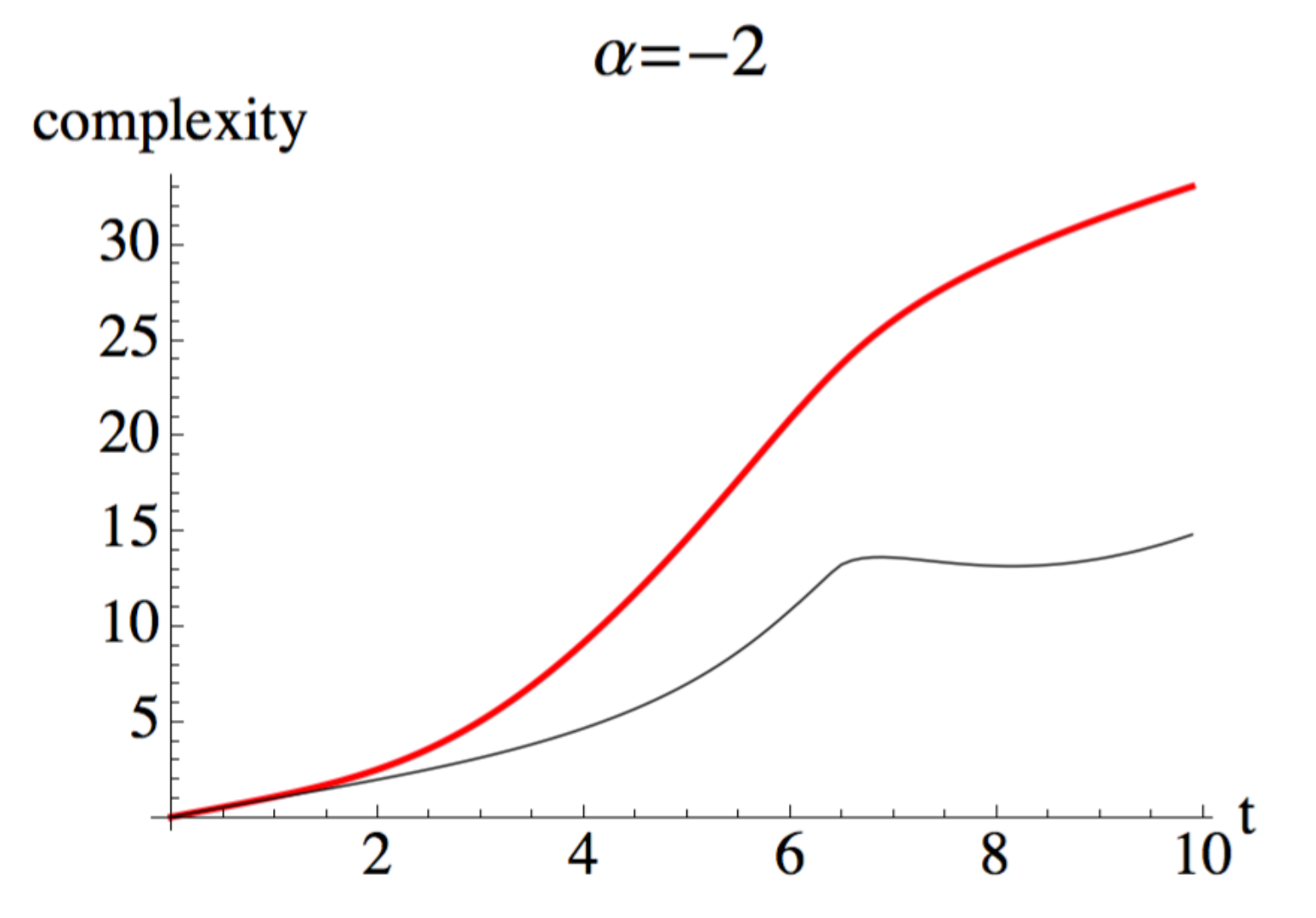}}
 \scalebox{0.40}{\includegraphics{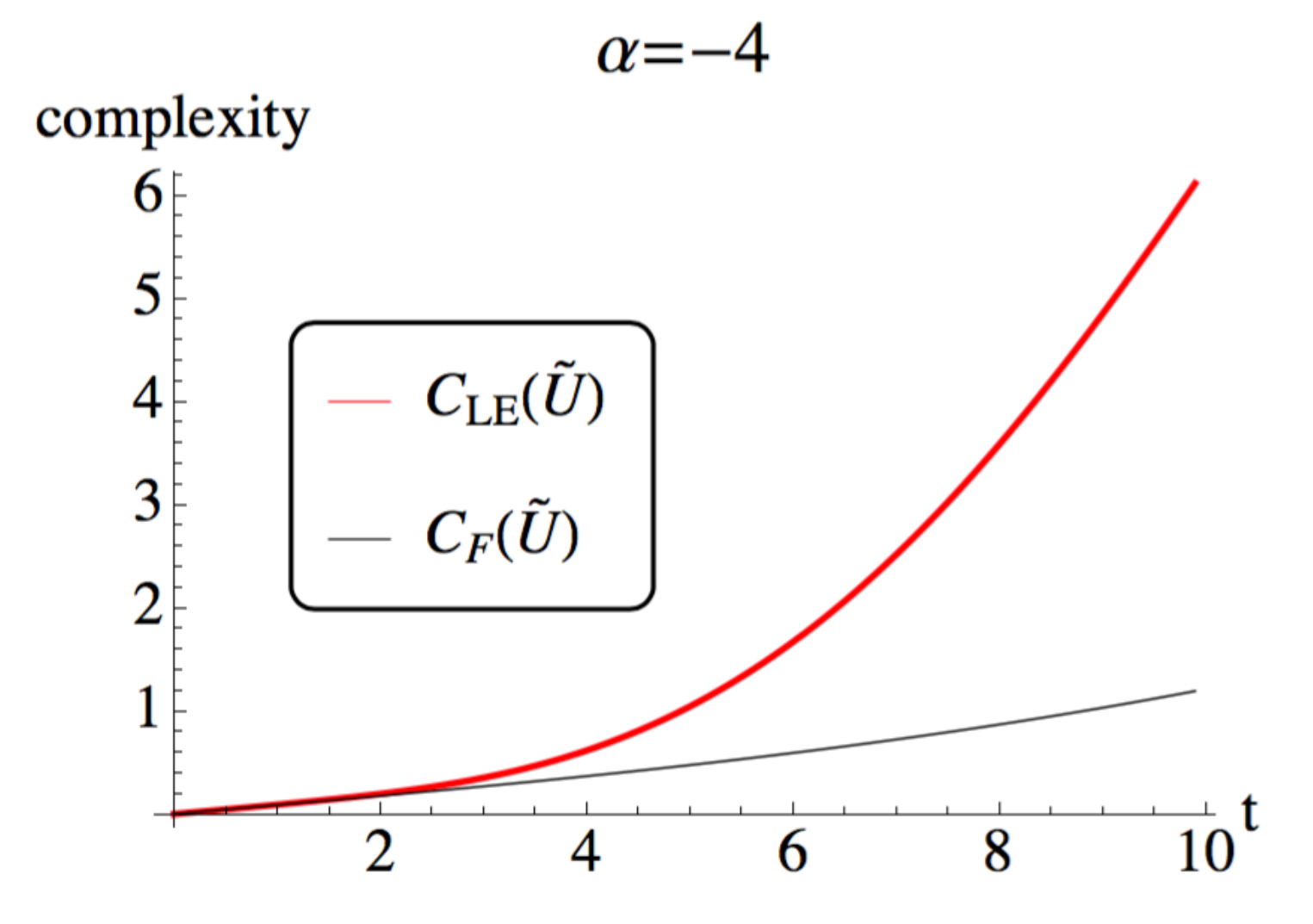}}
\caption{Comparison between early time growth  of circuit complexity for local and non-local theory}
\label{CC1}
\end{figure}
One might expect that this issue can be cured by considering interacting theories; here we take an alternative route -- we consider a Hamiltonian of the form in Eq.~(\ref{HBfourier}), 
but with a more general dispersion relation, namely
\be \label{dis}
\omega_k=\left(q^2+\hat q \cos\left(\frac{2\pi k}{N}\right)\right)^{\alpha/2}.
\ee
For $\alpha=1,$ this is the model considered in this work; here we consider (positive and negative) integer values of $\alpha$ with $|\alpha| > 1$ 
i.e. we consider a non-local theory. 
Fig.~(\ref{CC1}) shows the early time ($t < 10$) behavior of the circuit complexity for both $\mathcal{C}_{\text{LE}}(\tilde U)$ and $\mathcal{C}_{\text{F}}(\tilde U)$ for several values of 
$\alpha$.\footnote{We have set $N=500$ and used the values of the parameters in Eq.~(\ref{paravalue}).} 
Notice that the complexity grows for a substantial amount of time in these non-local theories i.e. we get the desired time-dependence of the complexity; the more non-local the theory is, the slower the rate of growth. 
[Also notice that the difference between $\mathcal{C}_{\text{LE}}(\tilde U)$ and $\mathcal{C}_{\text{F}}(\tilde U)$ becomes more pronounced as the theory becomes more non-local.]
In non-local theories, the entanglement entropy exhibits volume-law scaling (compared to the area-law scaling exhibited by local theories) \cite{TAL}; we speculate the volume-law (or area-law) scaling of the entanglement entropy is related to the growth of the complexity.  We investigate this in detail in a forthcoming paper \cite{WP}.



\section{Discussion}

In this work, we proposed a litmus test, titled {\bf LE vs F Test}, to distinguish between different methods of computing complexities. 
This test was predicated on the fact that an overlap between two states is invariant under unitary evolution; we categorized the different overlaps as Loschmidt echo and fidelity and computed the complexity between the states involved in the overlap. The idea was to investigate if the different measures of complexity are sensitive enough to capture the evolution of states; intriguingly we found that only circuit complexity from the Nielsen approach passes the test. This lead us to conclude that circuit complexity from the Nielsen approach is a more sensitive method, at least from the perspective of sensing time evolution of states.
Finally, we examined the nature of the growth of complexity for our model---the complexity grows very quickly, saturates, and then exhibits fluctuations around the saturated value. 
We observed that if we make the theory non-local (namely, considered a dispersion relation of the form in Eq.~\ref{dis} with $|\alpha| >1$), the complexity grows more slowly. \par

There are many interesting directions to pursue. It is important to understand the issue of the time scales, namely the relation between the equilibration time and the time for the complexity to saturate.
Until now, we have worked primarily in a discretized set-up; we would like take a continuum limit to make better contact with (continuous) QFTs. 
Moreover, to make contact with holography, it is important to generalize our results to interacting theories; toward that goal, one starting point could be to use the construction of \cite{Ame}. 
It is known that the Loschmidt echo can be used as a diagnostic for chaos, and one can extract information about the Lyapunov exponent from it \cite{Echo, Echothesis}; 
it would be interesting to evaluate $\mathcal{C}_{\text{LE}}(t)$ for chaotic theories and study its relation to chaos. 
Furthermore, it would be valuable to extend this construction to mixed states, which would create a platform to test some of the holographic results related to sub-region complexity \cite{Agon}.
Last but not the least, tensor networks provide useful ways to represent the time evolution of wave functions \cite{Milsted}; it would be interesting to understand if this kind of computation could shed light on the optimal network required to represent time evolution, thereby improving such constructions.  One could also study the causal structure of spacetime \cite{causnet}, and understand the connection between our construction and various path integral approaches\cite{Taka, Vidal,Vidala,Vidalb}.


\section*{Acknowledgment}

\noindent
We would like to thank Pawel Caputa, Michal P. Heller, Ro Jefferson, Esperanza Lopez, Jose Juan Melgarejo, Hugo Marrochio, Rob Myers, Aninda Sinha, Joan Simons, Tadashi Takayanagi, Javier Molina Vilaplana  for useful discussions and insightful comments. SSH would like to thank Perimeter Institute, where part of the work was done. AB also thanks Department of Physics, University of Murcia, IFT Madrid, Galileo Galilee Institute, Florence,  University of Geneva and  Max Planck Institute for Gravitational Physics, Potsdam for generously hosting him during the course of this work. 
AB is supported by JSPS Grant-in-Aid for JSPS fellows (17F17023). 
EHK acknowledges support from funds from the University of Windsor.
NM is supported by the South African Research Chairs Initiative of the Department of Science and Technology and the National Research Foundation (NRF) of South Africa; 
he was supported, in part, by the Perimeter Institute for Theoretical Physics during the course of this work.
Research at Perimeter Institute is supported by the Government of Canada through the Department of Innovation, Science, and Economic Development, and by the Province of
Ontario through the Ministry of Research and Innovation. 
Any opinion, finding and conclusion or recommendation expressed in this material is that of the authors and the NRF does not accept any liability in this regard.


\appendix


\section{Time Evolution of the Ground State Wave Function}

In this Appendix, we provide the details for obtaining the time-evolved wave function (\ref{tarw}), starting from the ground state of the Hamiltonian (\ref{H}).

The ground state of \ref{H} is given by
\be
\psi(\tilde x_k,0)=\prod_{k=0}^{N-1}\mathcal{N}_{k}(t=0)\exp\Big(-\frac{1}{2}\,\omega_k\, \tilde x_k\Big) 
\ee
with $\mathcal{N}_{k}(t=0)=\Big(\frac{\omega_k}{\pi}\Big)^{1/4}.$
The time-evolved wave function (\ref{tarw}) is obtained via
\begin{align}
\begin{split} \label{intform}
\psi(\tilde x_k, t)=\prod_{k=0}^{N-1}\mathcal{N}_{k}(t=0) \int_{-\infty}^{\infty} d\tilde x'_{k} 
  K^k(\tilde x_{k}, t|\tilde x'_{k}, t=0)\exp \Big(-\frac{1}{2}\omega_{k}\tilde x'_{k}\Big) \ ,
\end{split}
\end{align}
where the kernel is
\be
K^k (\tilde x_{k}, t|\tilde x'_{k}, t=0)=\langle \tilde x_k| \exp(-i\, t\, H_1(q_1,\hat q_1,q'_1))|\tilde x'_k\rangle \ ,
\ee
with $H_1(q_1,\hat q_1,q'_1)$ defined in (\ref{H1}); explicitly,
\be
K^k(\tilde x_{k}, t|\tilde x'_{k}, t=0)=\Big(\frac{\omega_{1,k}}{i\,2\pi\,\sin(\omega_{1,k}\,t)}\Big)^{1/2}\exp\Big\{\frac{i\,\omega_{1,k}}{2}\Big[((\tilde x_k)^2+(\tilde x'_k)^2) \cot(\omega_{1,k}\,t)\Big]-\frac{2\,\tilde x_{k}\,\tilde x'_{k}}{\sin (\omega_{1,k}\, t)}\Big\} \ .
\ee
Carrying out the Gaussian integral(s), one obtains
\begin{align}
\begin{split}
\psi(\tilde x_k,t)=\prod_{k=0}^{N-1}\mathcal{N}_{k}(t)\exp\Big(-\frac{1}{2}\Omega_{k}\tilde x_{k}^2\Big) \ ,
\end{split}
\end{align}
where
\begin{equation}
\Omega_{k}=\omega_{1,k} \Big [\frac{\omega_{1,k}-i\,\omega_k \cot (\omega_{1,k}\, t)}{\omega_k-i\,\omega_{1,k}\cot (\omega_{1,k}\, t)}\Big],\,\, 
\mathcal{N}_k=\Big(\frac{\omega_k}{\pi}\Big)^{1/4}\Big[\frac{1}{\omega_k-i\,\omega_{1,k}\cot(\omega_{1,k}\, t)}\Big]^{1/2}
  \Big(\frac{\omega_{1,k}}{i\,\sin(\omega_{1,k}\,t)}\Big)^{1/2} \ .
\end{equation}


\section{Choice of the Reference State}

In the main text, we discussed the complexity associated with the two overlaps (\ref{LS}) and (\ref{F}). For that we computed the relative complexity between some time evolved state and the ground state of the Hamiltonian (\ref{H}). In general, given a target state, the value of the complexity depends on the choice of reference state.  A natural choice for the reference state is an unentangled state, namely a state which has no entanglement in the original coordinate basis; this unentangled reference state is the ground state of the ultra-local Hamiltonian
\begin{equation}
 H_0= \frac{1}{2} \sum_k \left[ \tilde  p_{k}\tilde p_{-k} 
    + \omega_0^2~ \tilde x_{k}\tilde x_{-k}\right] \ ,
\label{Hultra}
\end{equation}
where $\omega_0$=constant i.e. it is dispersionless.  
Here we compute the complexity w.r.t. such an unentangled state, for the sake of the comparison with results in the main text.

With respect to this state, one can readily write down the expression for the complexity as evaluated in Sections  (3.2) and (3.3) --- we replace $\omega_k$ by $\omega_0$ in (\ref{circ}) and (\ref{comCov}).
The complexity from the Fubini-Study approach, however, is more involved; in what follows, we outline the calculation.  As done in Section (2), one can diagonalize $H_0$ by introducing
\begin{equation}
 \tilde  x_k = \frac{1}{\sqrt{2\omega_0} } \left( 
   c^{\phantom \dagger}_k + c^{\dagger}_{-k} \right)
 \ \ , \ \ 
 \tilde p_k = \frac{1}{i} \sqrt{\frac{\omega_0}{2}}
  \left( c^{\phantom \dagger}_k - c^{\dagger}_{-k} 
  \right)  \ ;
\end{equation} 
one obtains
\be
H_0=\sum_{k=0}^{N-1}\omega^0 (c_k^{\dagger}c_{k}+\frac{1}{2}) \ .
\ee
Next the operators $\{a_{k},a^{\dagger}_{-k}\}$ of (\ref{H}) are related to $\{c_{k},c^{\dagger}_{-k}\}$ via
\begin{align}
\begin{split}
\left(
\begin{array}{c}
 c_{k}  \\
 c^{\dagger}_{-k}  \\
\end{array}
\right)=\left(
\begin{array}{cc}
 \mathcal{U}^0_k & \mathcal{V}^0_k \\
\mathcal{V}^0_k &  \mathcal{U}^0_k \\
\end{array}
\right) \left(
\begin{array}{c}
 a_{k}  \\
 a^{\dagger}_{-k}  \\
\end{array}
\right)  \ ,
\end{split}
\end{align}
where
\begin{align}
\begin{split}
\mathcal{U}^0_{k}=\frac{\omega^0+\omega_{k}}{2\sqrt{\omega^0\omega_{k}}}
\ \ , \ \ 
\mathcal{V}^0_{k}=\frac{\omega^0-\omega_{k}}{2\sqrt{\omega^0\omega_{k}}} \ , 
\end{split}
\end{align}
with $|\mathcal{U}^0_k|^2-|\mathcal{V}^0_k|^2=1$.
Then the ground state of $H_0$, which we denote by $|\psi_{r}\rangle$, is given in terms of ground state of (\ref{H1}) as
\be \label{unentref}
|\psi_r\rangle=\prod_{k=0}\frac{1}{(\mathcal{U}^0_k)^2}\exp\Big[\gamma^0_k\, a^{\dagger}_{k}a^{\dagger}_{-k}\Big] |k,-k\rangle,
\ee
where $|k,-k\rangle$ is the $\{ a_k \}$ (Fock) vacuum and $\gamma^0_k=-2\Big(\frac{\mathcal{V}_k^0}{\mathcal{U}^0_{k}}\Big)$. 
With respect to this state (\ref{unentref}), the complexity for the time-evolved state (\ref{evolv1}) is given by Eq.~(\ref{genFS}) with
\be
\theta_{1,k}=2\arctanh |\gamma_{k}^0|
\ \ , \ \ 
\theta_{2,k}=2\arctanh |\gamma_{1,k}| 
\ \ , \ \ 
\phi_{1,k}=\pi, \phi_{2,k}=\arccos \Big[\Re\Big(\frac{\gamma_{1,k}}{|\gamma_{1,k}|}\Big)\Big] \ .
\ee

\section{A perturbative proof of $\mathcal{C}_{LE}(\tilde U)=\mathcal{C}_F(\tilde U)$ for Fubini-Study and covariance matrix method} \label{AppC}
 Here we present a perturbative proof for the equivalence of the two expressions in (\ref{diff1}) and (\ref{diff2}).  We show below, $\mathcal{C}_{LE}(\tilde U)=\mathcal{C}_F(\tilde U),$ to the leading order in small time expansion.
 \subsection*{LE vs F test using Fubini-Study method :}
 From (\ref{def}) and (\ref{defa}), $\alpha_{i,k},\beta_{i,k},\mu_{i,k}$ for $i=1,2$ are of $\mathcal{O}(t)$, in fact linear in $t$. Now we expand all the expressions in small $t$ and keep only the leading order term. 
 $$\sinh(\mu_{i,k}) =\mu_{i,k}+\mathcal{O}(t^2)+\cdots,\frac{1} {\cosh(\mu_{i,k})-\frac{\beta_{i,k}}{2\mu_{i,k}}\sinh(\mu_{i,k})}= 1+\mathcal{O}(t)+\cdots, i=1,2.$$
  So from (\ref{def1}) and (\ref{defa}) we can easily approximate $\gamma_{i,k}$ for $i=1,2$ as follows,
 \be
 \gamma^{\pm}_{i,k}= \alpha_{i,k}+\mathcal{O}(t^2)+\cdots, i=1,2. \ee
Also,
\be
\hat \gamma_k=\gamma_{2,k}^{+}+\frac{\gamma_{1,k}^{+}\gamma_{2,k}^{0}}{1-\gamma_{1,k}^{+}\gamma_{2,k}^{-}}= \gamma_{1,k}+\gamma_{2,k}+\mathcal{O}(t^2)+\cdots=\alpha_{1,k}+\alpha_{2,k}+\mathcal{O}(t^2)+\cdots.
\ee
From this we get,
\be
\arctanh |\hat \gamma_k|= |\hat \gamma_k|= |\alpha_{1,k}+\alpha_{2,k}|+\mathcal{O}(t^2)+\cdots.
\ee
Next,
\begin{align}
\begin{split}
&\theta_{i,k}=2\,\arctanh |\gamma_{1,k}|= |\gamma_{i,k}|+\mathcal{O}(t^2)+\cdots, i=1,2.
\end{split}
\end{align}
Using this we make the following expansion,
\be
\Re\Big(\frac{\gamma_{1,k}}{\gamma_{2,k}}\frac{|\gamma_{2,k}|}{|\gamma_{1,k}|}\Big)=1+\mathcal{O}(t^2)+\cdots
\ee
and 
\be
\arcosh\Big[\cosh (\theta_{1,k})\cosh(\theta_{2,k})-\sinh(\theta_{1,k})\sinh(\theta_{2,k})\Re\Big(\frac{\gamma_{1,k}}{\gamma_{2,k}}\frac{|\gamma_{2,k}|}{|\gamma_{1,k}|}\Big)\Big]=  \frac{1}{2}|\alpha_{1,k}+\alpha_{2,k}|+\mathcal{O}(t^2)+\cdots.
\ee
Then from (\ref{diff1}) we can easily see that,
\begin{align}
\begin{split} \label{expp}
\mathcal{C}_{LE}(\tilde U)=\mathcal{C}_{F}(\tilde U)=& t\, \sqrt{\sum_{k=0}^{N-1} \Big(\omega_{2,k} \, \mathcal{\tilde U}_k\mathcal{\tilde V}_k-\omega_{1,k} \, \mathcal{U}_k\mathcal{V}_k\Big)^2}+\mathcal{O}(t^2)+\cdots,\\&
=\frac{t}{2}\, \sqrt{\sum_{k=0}^{N-1} \Big(\frac{\omega_{1,k}^2-\omega_{2,k}^2}{\omega_k}\Big)^2}+\mathcal{O}(t^2)+\cdots.
\end{split}
\end{align}
 
 \subsection*{LE vs F test  using covariance matrix  method:}
 
 From (\ref{smallt}), small $t$ expansion gives,
 \begin{align}
 \begin{split}
 &\Re(\Omega_k)=\omega_{k}+t^2\, \omega_{k}\, \left(\omega_{1,k}^2 -\omega_{k}^2\right)+\mathcal{O}(t^3)+\cdots,\\&
 \text{Im} (\Omega_k)= t\, \left(\omega_{1,k}^2 -\omega_{k}^2\right)+\mathcal{O}(t^2)+\cdots
 \end{split}
 \end{align}
 We obtain similar expressions for $\Omega_{1,k}$ with $\omega_{1,k}$ in the above expressions replaced by $\omega_{2,k}.$
 Then we get,
 \begin{align}
 \begin{split}
 &|\Omega_{1,k}|^2+|\Omega_k|^2-2 \text{Im}(\Omega_{1,k})\text{Im}(\Omega_k)\\&= 2\omega_k^2+t^2\Big((\omega_{1,k}^2-\omega_{2,k}^2)^2+2(\omega_{1,k}^2+\omega_{2,k}^2)\omega_k^2-4\omega_k^4\Big)+\mathcal{O}(t^3)+\cdots.
 \end{split}
 \end{align}
 Also we have,
 \be
 \frac{1}{2\Re(\Omega_{1,k})\Re(\Omega_{k})}=\frac{1}{2\omega_k^2}-t^2\,\Big( \frac{\omega_{1,k}^2+\omega_{2,k}^2-2\omega_k^2}{2\omega_k^2}\Big)+\mathcal{O}(t^3)+\cdots.
 \ee
 Putting all these together we finally get,
 \be
\Big( \arcosh \left(\frac{\Re(\Omega_{1,k})^2+\Re(\Omega_k)^2+(\text{Im}(\Omega_{1,k})-\text{Im}(\Omega_k))^2}{2\,\Re(\Omega_{1,k}) ,\Re(\Omega_k)}\right)\Big)^2=t^2\frac{(\omega_{1,k}^2-\omega_{2,k}^2)^2}{\omega_k^2}+\mathcal{O}(t^3)+\cdots.
 \ee
 On the other hand from (\ref{omedef}) we can easily see that,
 \begin{align}
 \begin{split}
 \Re(\hat \Omega_k)=\omega_k+ t^2\,\omega_k\,  \left(\omega_{2,k}^2-\omega_{1,k}^2\right)+\mathcal{O}(t^3)+\cdots,
\text{Im} (\hat \Omega_k)=t\, \left(\omega_{1,k}^2-\omega_{2,k}^2\right)+\mathcal{O}(t^2)+\cdots.
 \end{split}
 \end{align}
 Using this it's not hard to show that,
 \be
\left( \arcosh \left( \frac{\omega_k^2+ |\hat \Omega_k |^2 }{2 \  \omega_k \Re (\hat \Omega_k)}\right) \right)^2=t^2\frac{(\omega_{1,k}^2-\omega_{2,k}^2)^2}{\omega_k^2}+\mathcal{O}(t^3)+\cdots.
 \ee
 Then from (\ref{diff2}) we get,
 \be   \label{expp1}
\mathcal{C}_{LE}(\tilde U)=\mathcal{C}_{F}(\tilde U)= \frac{t}{2}\, \sqrt{\sum_{k=0}^{N-1} \Big(\frac{\omega_{1,k}^2-\omega_{2,k}^2}{\omega_k}\Big)^2}+\mathcal{O}(t^2)+\cdots
 \ee
 This proves our claim. We leave the  more general proof  (non-perturbative in $t$)  for a future investigation.  Interestingly we observe that at this order  expressions mentioned in (\ref{expp}) and (\ref{expp1}) are equal to each other.


\end{document}